\documentclass[final,5p,times]{elsarticle}

\usepackage{amssymb}
\usepackage{amsmath}
\usepackage{physics}
\usepackage{hyperref}
\usepackage{braket}
\usepackage{longtable}
\usepackage{multirow}
\usepackage{listings}
\usepackage{xcolor}

\definecolor{codegreen}{rgb}{0,0.6,0}
\definecolor{codegray}{rgb}{0.5,0.5,0.5}
\definecolor{codepurple}{rgb}{0.58,0,0.82}
\definecolor{backcolour}{rgb}{0.95,0.95,0.92}

\lstdefinestyle{mystyle}{
  backgroundcolor=\color{backcolour},   commentstyle=\color{codegreen},
  keywordstyle=\color{magenta},
  numberstyle=\tiny\color{codegray},
  stringstyle=\color{codepurple},
  basicstyle=\ttfamily\footnotesize,
  breakatwhitespace=false,         
  breaklines=true,                 
  captionpos=b,                    
  keepspaces=true,                 
  numbers=left,                    
  numbersep=5pt,                  
  showspaces=false,                
  showstringspaces=false,
  showtabs=false,                  
  tabsize=2
}

\newcounter{bla}

\journal{Computer Physics Communications}
\newcommand{\comment}[1]{}

\graphicspath{ {./figures/} }

\begin{document}

\begin{frontmatter}
\title{{\tt dmscatter}: A Fast Program for WIMP-Nucleus Scattering} 

\author[add1]{Oliver C. Gorton}\ead{ogorton@sdsu.edu}
\author[add1]{Calvin W. Johnson}\ead{cjohnson@sdsu.edu}
\author[add2,add1]{Changfeng Jiao}\ead{jiaochf@mail.sysu.edu.cn}
\author[add3]{Jonathan Nikoleyczik}

\address[add1]{San Diego State University, San Diego, California 92182, USA}
\address[add2]{Sun Yat-sen University, Zhuhai 519082, China}
\address[add3]{University of Wisconsin, Madison, Wisconsin,
53706, USA}

\begin{abstract}
Recent work~\cite{Fitzpatrick_2013,Anand_2014}, using an effective field theory framework, have shown the number of possible couplings between nucleons and the dark-matter-candidate Weakly Interacting Massive Particles (WIMPs) is larger than previously thought. Inspired by an existing Mathematica script that computes the target response~\cite{Anand_2014}, we have developed a fast, modern Fortran code, including optional OpenMP parallelization, along with a user-friendly Python wrapper, to swiftly and efficiently explore many scenarios, with output aligned with practices of current dark matter searches.   A library of  most of the important target nuclides is included; users may also import their own nuclear structure data, in the form of reduced one-body density matrices.
The main output is the differential event rate 
as a function of recoil energy, needed for 
modeling detector response rates, 
but intermediate results such as 
nuclear form factors can be readily 
accessed.
\end{abstract}


\end{frontmatter}

{\bf PROGRAM SUMMARY}
\begin{small}
\noindent
{\em Program Title: dmscatter}                                          \\
{\em CPC Library link to program files:} (to be added by Technical Editor) \\
{\em Developer's repository link:} \href{https://github.com/ogorton/dmscatter}{github.com/ogorton/dmscatter}  \\
{\em Code Ocean capsule:} (to be added by Technical Editor)\\
{\em Licensing provisions:} MIT\\
{\em Programming languages: Fortran, Python}                                   \\
{\em Supplementary material:}                                 \\
{\em Nature of problem:} Simulating the event rate of nuclear recoils from collisions with 
dark matter for a variety of different nuclear targets and different nucleon-dark matter couplings.  \\
{\em Solution method:} The event rate is an integral over the product 
of the nuclear and dark matter response functions, weighted by the expected dark matter flux.
To compute the nuclear response, 
reduced one-body nuclear density matrices 
are coupled to multipole expansion of operators computed analytically in a harmonic oscillator 
basis. The nuclear structure input, in the form of one-body densities matrices, are supplied by 
outside code; we provide a library of the most prominent targets. 
\\
{\em Additional comments including restrictions and unusual features:} None.\\
   \\
\end{small}

\allowdisplaybreaks

\section{Introduction}

Substantial experimental effort has been and continues to be expended to directly detect  `dark matter,' some as-yet unidentified 
nonbaryonic particles which astrophysical and cosmological evidence suggests may make up a substantial fraction (roughly a quarter) of the universe's mass-energy~\cite{Blumenthal:1984bp,Bertone:2016nfn}.
Because dark matter interacts with baryonic matter weakly and may be very massive compared to baryonic particles--WIMPs or \textit{weakly interacting massive particles}--these experiments attempt to measure the recoil of nuclei from unseen and (mostly) elastic collisions~\cite{Primack:1988zm,Feng:2010gw}. 

Originally it was assumed  that dark matter particles would simply couple  either to the scalar or spin densities of  nucleons~\cite{PhysRevD.31.3059,RevModPhys.71.S197}.
But a few years ago Fitzpatrick \textit{et al.} used effective field theory (EFT) calculations assuming Galilean invariance to identify upwards of 15 possible independent couplings between nonrelativistic dark matter and  nucleons~\cite{Fitzpatrick_2013,Anand_2014,Vietze:2014vsa,Hoferichter:2020osn}. 

 The enlargement of the possible couplings motivates a variety of nuclear targets. More targets better constrain the 
 actual coupling, but also complicate simulations of detector responses. To aid such simulations, Anand, Fitzpatrick, and Haxton 
 made available a script written in Mathematica computing the dark matter event rate spectra~\cite{Anand_2014};  this script, \texttt{dmformfactor}, embodied the nuclear structure in a shell model framework and the user could choose the coupling to the EFT-derived operators. An updated Mathematica script has been developed and applied~\cite{HaxtonNewCode,xia2019pandax,DMreview}.
 
Not only did the EFT framework break new ground in the planning and analysis of dark matter direct detection experiments, the Mathematica script made the new framework widely accessible.  Like many scripts in interpreted languages, however, \texttt{dmformfactor} is not fast, and scanning through a large set of parameters, such as exploring the effects of mixing two or more couplings or carrying out uncertainty quantification~\cite{PhysRevC.101.054308}, ends up being time-consuming.

Inspired by the Mathematica script \texttt{dmformfactor}, we present here  a fast modern Fortran code, {\tt dmscatter}, for computing WIMP-nucleus scattering event rates using the previously proposed theoretical framework. The output is designed to align with practices of current dark matter searches.  With advanced algorithmic and numerical implementation, including the ability to take advantage of multi-core CPUs, our code opens up new areas of research: to rapidly explore the EFT parameter space including interference terms, and to conduct sensitivity studies to address the uncertainty introduced by the underlying nuclear physics models. Furthermore, we enhance the accessibility by including Python wrapper and example scripts and which can be used to call the Fortran code from within a Python environment.

(Similar to {\tt dmscatter} is the DDCalc 
dark matter direct detection phenomenology package~\cite{bringmann2017darkbit,athron2019global}. Both DDCalc and 
{\tt dmscatter} use efficient Fortran90 central engines with Python 
wrappers and allow for the full panoply of nonrelativistic couplings. 
While DDCalc specifically incorporates tools to predict  signal rates and likelihoods for ongoing 
dark matter experiments,  {\tt dmscatter} has more flexible 
 nuclear structure input, allowing the exploration and propagation of the 
uncertainty in the nuclear physics.)
\section{Theoretical Background}

Although the formalism is fully developed and presented in the original papers ~\cite{Fitzpatrick_2013,Anand_2014,Vietze:2014vsa,Hoferichter:2020osn}, for completeness and convenience we summarize the main ideas here. 

\subsection{Differential event rate}

The key product of the code is the differential event rate for WIMP-nucleus scattering events in number of events per MeV. This is obtained by integrating the differential WIMP-nucleus cross section over the velocity distribution of the WIMP-halo in the galactic frame:
\begin{equation}\label{ER}
	\frac{dR}{dE_r}(E_r)
	 = N_T n_\chi \int \frac{d\sigma}{dE_r}(v,E_r)\ \tilde{f}(\vec{v})\ v\ d^3v,
\end{equation}
where $E_r$ is the recoil energy of the WIMP-nucleus scattering event, $N_T$ is the number of target nuclei, $n_\chi = \rho_\chi/m_\chi$ is the local dark matter number density, $\sigma$ is the WIMP-nucleus cross section.  The dark matter velocity distribution in the lab frame, $\tilde{f}(\vec{v})$, is obtained by boosting the Galactic-frame distribution $f(\vec{v})$: $\tilde{f}(\vec{v}) = f(\vec{v} + \vec{v}_{earth})$, where $\vec{v}_{earth}$ is the velocity of the earth in the galactic rest frame. 

\subsection{Halo model}

There are many models for the dark matter distributions of galaxies. We provide the Simple Halo Model (SHM) with smoothing, a truncated three-dimensional Maxwell-Boltzmann distribution:
\begin{equation}\label{eq: f}
    f(\vec{v}) = \frac{\Theta(v_{esc}-|\vec{v}|)}
        {N_{esc}\pi^{3/2}v_0^3}
        \left\{\exp[-(\vec{v}/v_0)^2] - \exp[-(v_{esc}/v_0)^2] \right\},
\end{equation}
where $v_0$ is some scaling factor (typically taken to be around $220\ \mathrm{km/s}$),
and $N_{esc}$ re-normalizes due to the cutoff \cite{PhysRevD.33.3495, PhysRevD.37.3388}.
Halo distributions are not the focus of this paper, and we leave the implementation of more sophisticated halo models, such as SMH++ \cite{evans2018shm}, to future work.
The integral in equation (\ref{ER}) is evaluated numerically. Details can be found in \ref{sec: integral}.

\subsection{Differential cross section}

The differential scattering cross section is directly related to the scattering transition probabilities $T(v,q(E_r))$:
\begin{align}
    \frac{d\sigma}{dE_r}(v,E_r) = 2m_t\frac{d\sigma}{dq^2}(v,q) = \frac{2m_T}{4\pi v^2} T(v,q).
\end{align}
The momentum transfer $q$ is directly related to the recoil energy by $q^2=2m_tE_r$, where $m_t$ is the mass of the target nucleus in GeV$/c^2$.

\subsection{Transition probability}

The WIMP-nucleus scattering event probabilities are computed as a sum of squared nuclear-matrix-elements:
\begin{multline}\label{eq:T1}
    T(v,q) = \frac{1}{2j_\chi +1}\frac{1}{2j_T+1} \sum_{M_i M_f} \left |\Braket{j_TM_f | \mathcal{H} | j_TM_i}\right |^2 \\
    = \frac{1}{2j_\chi +1}\frac{1}{2j_T+1} \left |\Braket{j_T || \mathcal{H} || j_T}\right |^2 
\end{multline}
which we have rewritten in terms of reduced (via the Wigner-Eckart theorem) matrix elements~\cite{edmonds1996angular}, as denoted by the double bars $||$. 
Here $v$ is the speed of the WIMP in the lab frame, $q$ is the momentum transferred in the collision, and $j_\chi$ and $j_T$ are the intrinsic spins of the WIMP and target nucleus, respectively. $\mathcal{H}$ is the WIMP-nucleus interaction.

\subsection{WIMP-nucleus interaction}

The WIMP-nucleus interaction $\mathcal{H}$ is defined in terms of the effective field theory Lagrangian constructed from all leading order combinations of the following operators:
\begin{equation}
    i\frac{\vec{q}}{m_N},\ \vec{v}^\perp,\ \vec{S}_\chi,\ \vec{S}_N.
\end{equation}
$\vec{v}^\perp$ is the relative WIMP-target velocity and $\vec{S}_\chi, \vec{S}_N$ are the WIMP and nucleon spins, respectively. There are fifteen such combinations, listed in \ref{sec: EFT}, and $\mathcal{H}$ is specified implicitly by corresponding coupling constants $c_i^x$, (for $i=1,...,15$), where $x=p, n$ for coupling to protons or neutrons individually:
\begin{equation}
    \mathcal{H} = \sum_{x=p,n}\sum_{i=1,15} c^x_i \mathcal{O}_i^x
\end{equation}

\subsection{Coupling coefficients}

The EFT coefficients $c_i^x$ can be expressed in terms of proton and neutron couplings $x=p,n$, or, equivalently, isospin and isovector couplings $\tau=0,1$. The relationship between the two is:
\begin{align}
    c^{\tau=0}_i &= \frac{1}{2}(c^{x=p} + c^{x=n}_i) \\
    c^{\tau=1}_i &= \frac{1}{2}(c^{x=p} - c^{x=n}_i).
\end{align}
Our code accepts either specification and automatically converts between the two. {(Note: 
 while \cite{Anand_2014} specifies this same relationship, the Mathematica script 
distributed along with it in Supplementary Material, {\tt dmform\-factor-prc.m}, also distributed as  {\tt dmformfactor-V6.m}, 
actually uses a different transformation, namely  
$    c^{\tau=0,1}_i = c^{x=p}_i \pm c^{x=n}_i.$ 
Later versions~\cite{HaxtonNewCode,xia2019pandax,DMreview} however, are consistent with the 
above relationship.)}

\subsection{Response functions}

The summand of equation (\ref{eq:T1}) is ultimately factorized into two factors: one containing the EFT content, labeled $R_i^{x,x'}$, and another containing the nuclear response functions, labeled $W_i^{x,x'}$ for each of the $i=1,...,8$ allowed combinations of electro-weak-theory, discussed in the next section. The former are listed in \ref{sec: wimpresponse}.

There are eight nuclear response functions $W_i^{x,x'}$ considered here. The first six nuclear response functions have the following form:
\begin{equation}
\label{responsefunctions}
    W_{X}^{x,x'} = \sum_{J}\bra{\Psi} | X^{x}_J | \ket{\Psi}\bra{\Psi} | X^{x'}_J |\ket{\Psi},
\end{equation}
with $X$ selecting one of the six electroweak operators:
\begin{equation}
    X_J=M_J, \Delta_J, \Sigma_J', \Sigma_J'', \tilde{\Phi}_J', \Phi_J''.
\end{equation} 
(further described in section \ref{sec: ops}) and $\Psi$ being the nuclear wave function for the ground state of the target nucleus. 
The sum over operators spins $J$ is restricted to even or odd values of $J$, depending on restrictions from conservation of parity and charge conjugation parity (CP) symmetry.

As a check of normalization, the $J=0$  contribution to $W^{xx^\prime}_{M}$ is just the square of the 
Fourier transform of the rotationally invariant density. (For even-even targets, this is the 
only contribution to ground state densities.)  This means, at momentum transfer $q=0$, the 
$J=0$ contribution to 
$W^{pp}_{M} = Z^2/4\pi$, $W^{nn}_{M} = N^2/4\pi$, and for isospin-format form factors, the 
$J=0$ contribution to 
$W^{00}_{M} = \frac{1}{4\pi} \left (\frac{A}{2} \right )^2$. Such limits are useful when comparing 
to other calculations, to ensure agreement in normalizations.

Two additional response functions add interference-terms:
\begin{equation}
    W_{M\Phi''}^{x,x'} = 
    \sum_{J}\bra{\Psi} | M_{J}^{x} |\ket{\Psi}\bra{\Psi} |\Phi_{J}^{''x'} |\ket{\Psi}
    ,
\end{equation}   
\begin{equation}
    W_{\Delta\Sigma'}^{x,x'} = \sum_{J}\bra{\Psi} |\Sigma_{J}^{'x} |\ket{\Psi}\bra{\Psi}  |\Delta_{J}^{x'} | \ket{\Psi} .
\end{equation}
The transition probability is thus \cite{Fitzpatrick_2013,Anand_2014}:
\begin{multline}\label{eq:T2}
    T(v,q) =\frac{4\pi}{2j_T+1} \sum_{x=p,n}\sum_{x'=p,n}\sum_{i=1}^8 R_i^{x,x'}(v^2,q^2)W_i^{x,x'}(q),
\end{multline}
where $i\to X$ for $i=1,..,6$, and $i=7 \to M\Phi''$, $i=8\to \Delta\Sigma'$.

\subsection{Nuclear operators}\label{sec: ops}

There are six basic 
 operators, $M_J, \Delta_J, \Sigma_J', \Sigma_J'', \tilde{\Phi}_J', \Phi_J''$, describing the electro-weak coupling of the WIMPs to the nucleon degrees of freedom.
These are constructed from Bessel spherical and vector harmonics \cite{DONNELLY1979103}:
\begin{align}
    M_{JM}(q\vec{x})&\equiv j_J(qx)Y_{JM}(\Omega_x)\\
    \vec{M}_{JML}(q\vec{x}) &\equiv j_L(qx) \vec{Y}_{JLM}(\Omega_x),
\end{align}
where, using unit vectors $\vec{e}_{\lambda = -1, 0, +1}$, 
\begin{align}
    Y_{JLM}(\Omega_x) = \sum_{m\lambda} \bra{Lm1\lambda}\ket{(L1)JM_J} Y_{Lm}(\Omega_x)\vec{e}_\lambda.
\end{align}
The six multipole operators are defined as:
\begin{align}
    \label{oplist}
M_{JM}\ \ &\\
\Delta_{JM} \equiv& \vec{M}_{JJM}\cdot \frac{1}{q}\vec{\nabla}\\
\Sigma'_{JM} \equiv& -i \left \{\frac{1}{q}\vec{\nabla}\times \vec{M}_{JJM}  \right\}\cdot \vec{\sigma}\\
\Sigma''_{JM} \equiv& \left \{ \frac{1}{q}\vec{\nabla}M_{JM} \right \}\cdot \vec{\sigma}\\
\tilde{\Phi}'_{JM} \equiv& \left( \frac{1}{q} \vec{\nabla} \times \vec{M}_{JJM}\right)\cdot \left(\vec{\sigma}\times \frac{1}{q}\vec{\nabla} \right) + \frac{1}{2}\vec{M}_{JJM}\cdot \vec{\sigma}\\
\Phi''_{JM}\equiv& i\left(\frac{1}{q}\vec{\nabla}M_{JM} \right)\cdot \left(\vec{\sigma}\times \frac{1}{q}\vec{\nabla} \right)
\end{align}
The matrix elements of these operators can be calculated for standard wave functions from second-quantized shell model calculations:
\begin{equation}
    \bra{\Psi_f} | X_J| \ket{\Psi_i} 
 = \sum_{a,b}
     \bra{a} |X_J| \ket{b} 
     \rho^{fi}_J(ab),
\end{equation}
where single-particle orbital labels $a$ imply shell model quantum number $n_a, l_a, j_a$, 
and the double-bar $||$ indicates reduced matrix elements~\cite{edmonds1996angular}. For elastic collisions, only the ground state is involved, i.e. $\Psi_f=\Psi_i=\Psi_{g.s.}$.

\subsection{Nuclear structure}

We assume a harmonic oscillator single-particle basis, with the important convention that the 
radial nodal quantum number $n_a$ starts at 0, that is, we label the orbitals as $0s, 0p, 1s0d$, etc.., 
and \textit{not} starting with $1s, 1p,$ etc. 
By default, the harmonic oscillator basis length $b=\sqrt{\hbar/(m\omega)}$ is set to the Blomqvist and Molinari prescription \cite{BLOMQVIST1968545}:
\begin{align}\label{bfm}
    b^2 = 41.467/( 45A^{-1/3} - 25A^{-2/3})\ \mathrm{fm}^2.
\end{align}
Other values can be set using the control words \texttt{hofrequency} or \texttt{hoparameter} (see \ref{sec: cwords}).
Then, the one-body matrix elements for operators $ \bra{a} |X^{(f)}_J| \ket{b}$, built from spherical Bessel functions and vector spherical harmonics,  have closed-form expressions in terms of confluent hypergeometric 
functions~\cite{DONNELLY1979103}.

The nuclear structure input is in the form of one-body density matrices between many-body eigenstates,
\begin{equation}
\rho^{fi}_J(ab) = \frac{1}{\sqrt{2J+1} }\langle \Psi_f || [ \hat{c}^\dagger_a \otimes \tilde{c}_b ]_J 
|| \Psi_i \rangle, \label{eqn:denmat}
\end{equation}
where $\hat{c}^\dagger_a$ is the fermion creation operator (with good angular momentum quantum numbers), 
$\tilde{c}_b$ is the time-reversed~\cite{edmonds1996angular} fermion destruction operator.  Here the matrix element is reduced in angular momentum but not isospin, and so are in proton-neutron format. These density matrices are the 
product of a many-body code, in our case {\sc  Bigstick}~\cite{BIGSTICK1,BIGSTICK2}, although one could use one-body density matrices, appropriately formatted (see \ref{nuclides}), from any many-body code.

\section{Description of the code}

\begin{figure*}[ht!]
    \centering
    \includegraphics[width=.5\textwidth]{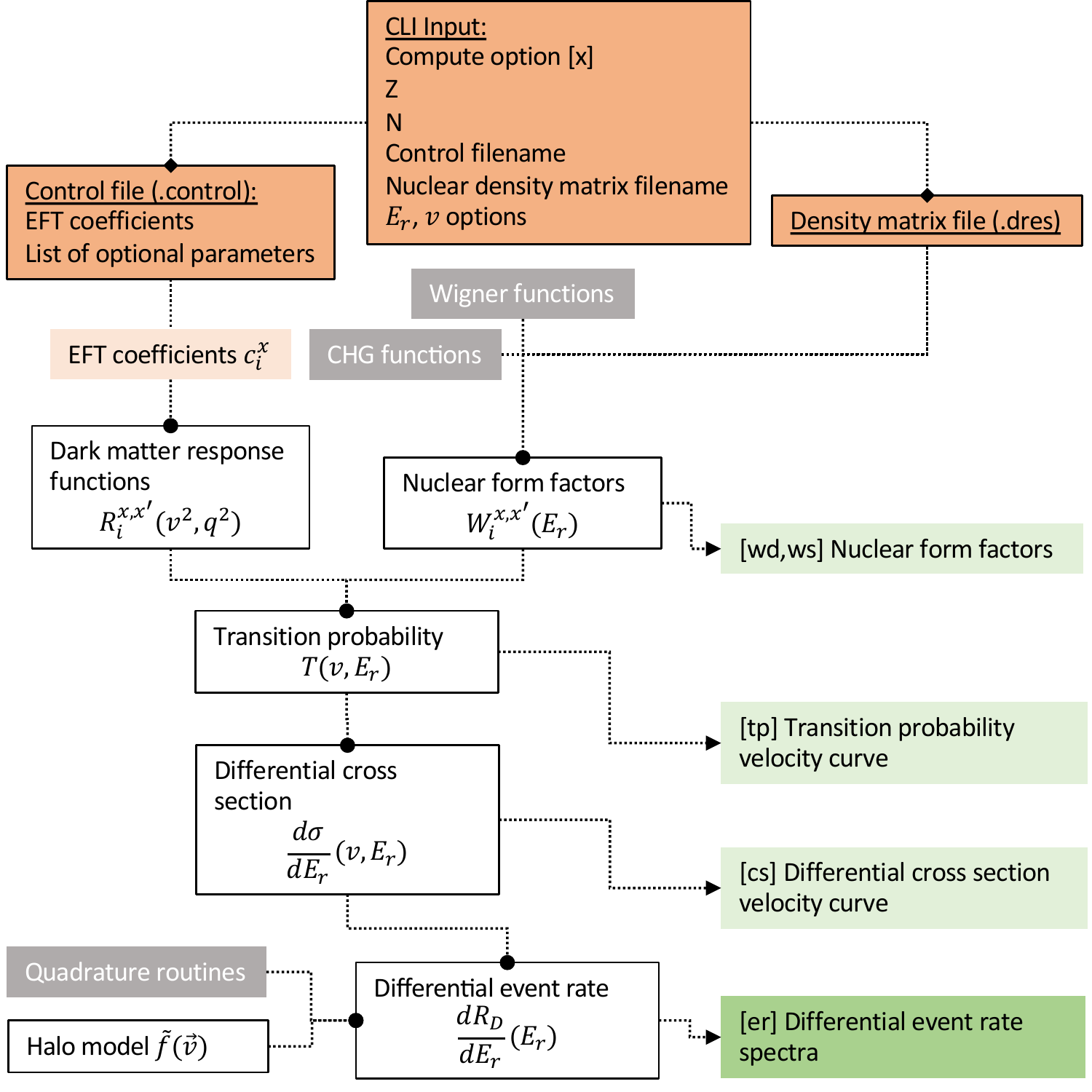}
    \caption{Flow of {\tt dmscatter}. Information at various steps can be extracted and compared to other calculations. Orange boxes represent input files. Green boxes represent quantities the code can write to file. White boxes are important steps of interest and grey boxes are libraries built into the code. There are three groups of inputs for a WIMP-nucleus scattering event calculation. (1) The nuclear target information:  target nucleus mass, spin, and one-body density matrix elements. (2) 
    The generic, non-relativistic EFT-specification of the WIMP-nucleon interaction,  in the form of coupling parameters along with the WIMP mass and spin. (3) The velocity distribution describing the relative motion of incoming WIMP particles in the laboratory frame. }
    \label{fig:flow_chart}
\end{figure*}

The structure of the code and its inputs are outlined in Figure \ref{fig:flow_chart}. The Fortran code replicates the capabilities 
of the { earlier Mathematica script}~\cite{Anand_2014}. 
Notably, one can compute the differential WIMP-nucleon scattering event rate for a range of recoil energies or transfer momenta, and any quantity required to determine those, such as the tabulated nuclear response functions. 

We provide a detailed manual as part of the distribution package, and a `quick start' guide can be found in section~\ref{running}. 
The central engine, \texttt{dmscatter}, is written in standard modern Fortran and has OpenMP for an easy and optional parallel speed-up. 
While the distributed Makefile assumes the GNU Fortran compiler \texttt{gfortran}, the code should be able to be compiled by most recent Fortran compilers and does not require any special compilation flags, aside from standard (and optional) optimization and parallel OpenMP  flags. 

We supply a library of nuclear structure files (one-body density matrix files) for many of the common expected targets, as listed in Table \ref{tab:includednuclei}.  (We also include, for purposes of validation, the legacy density matrices included with the original 
Mathematica script ~\cite{Anand_2014}.)
These density matrix files are written in plain ASCII, 
using the format output by the nuclear configuration-interaction code {\sc Bigstick}~\cite{BIGSTICK1,BIGSTICK2}.  The only assumption is 
that the single-particle basis states are harmonic oscillator states; the user must supply the harmonic oscillator single particle basis 
frequency $\Omega$, typically given in MeV as $\hbar \Omega$, or the related length parameter $b= \sqrt{\hbar/M\Omega}$,
where $M$ is the nucleon mass. 
\begin{table}[ht]
    \centering
    \begin{tabular}{l | p{3cm} | c}
    \hline
    \hline
        Nuclei & Isotopes & Source \\
        \hline\hline
        He & 4 & \cite{PhysRevC.68.041001,shirokov2016n3lo}\\
        C  & 12 &  \cite{cohen1965effective,PhysRevC.68.041001,shirokov2016n3lo}\\
        F & 19 & \cite{wildenthal1984empirical,brown1988status},\cite{PhysRevC.74.034315}\\
        Na & 23 & \cite{wildenthal1984empirical,brown1988status},\cite{PhysRevC.74.034315}\\
        Si & 28, 29 & \cite{wildenthal1984empirical,brown1988status},\cite{PhysRevC.74.034315}\\
        Ar & 40 &  \cite{PhysRevC.86.051301}\\
        Ge & 70, 72, 73, 74, 76 & \cite{PhysRevC.80.064323} \\
        I & 127 & \cite{PhysRevC.71.044317}; \cite{GCN5082} used in \cite{PhysRevLett.100.052503,PhysRevC.82.064304}\\ 
        Xe & 128, 129, 130, 131, 132, 134, 136 & \cite{PhysRevC.71.044317}; \cite{GCN5082} used in \cite{PhysRevLett.100.052503,PhysRevC.82.064304}\\
    \end{tabular}
    \caption{Table of nuclear data for targets we include with the program at time of publication. Each corresponds to a (.dres) density matrix file in the \texttt{targets} directory. The source indicates the nuclear Hamiltonian that was used to generate the wave function data. See the manual and GitHub repository 
    for updates and full information on provenance. New targets may be added in future releases.}
    \label{tab:includednuclei}
\end{table}

\section{Compiling and running the code}\label{running}

We provide a detailed manual for running the code as a separate document with the code distribution, available on our public GitHub repo: \href{https://github.com/ogorton/dmscatter}{github.com/ogorton/dmscatter}. To get started, one needs a modern Fortran  
compiler, the \texttt{make} tool, and, optionally, a Python interpreter (with NumPy, Matplotlib). We used the widely available GNU Fortran
(\texttt{gfortran}) compiler, but we use only standard Fortran and our code should be 
able to be compiled by other Fortran compilers as well; the user will have to modify the makefile.

\subsection{Compiling the code}

To compile {\tt dmscatter}, navigate to the \texttt{build} directory and run:
\begin{verbatim}
make dmscatter
\end{verbatim}
This will compile the code using \texttt{gfortran}, creating the executable \texttt{dmscatter} in the \texttt{bin} directory. If you want to use a different compiler, you must edit the corresponding Makefile line to set ``\texttt{FC = gfortran}'' e.g., by replacing \texttt{gfortran} with \texttt{ifort}.
To compile an OpenMP-parallel version of the code, use the option:
\begin{verbatim}
make openmp
\end{verbatim}
Note that if you switch between and serial or parallel version, you must first run \texttt{make clean}.

\subsection{Running the code}\label{runrunrun}

The command line interface (CLI) to the code prompts the user for:
\begin{enumerate}
    \item Type of calculation desired
    \item Number of protons $Z$ in the target nucleus 
    \item Number of neutrons $N$ in the target nucleus
    \item Control file name (.control)
    \item Nuclear structure file name (.dres)
    \item Secondary options related to the type of calculation
\end{enumerate}
For details on each type of input file, see sections~\ref{sec: cfile} and~\ref{nuclides}. Additionally, if the user enables the option ``usenergyfile'', then a file containing the input energies or momentum will also be required.
The secondary options are typically three numbers specifying the range of recoil energies or momentum for which to compute the output. After running, the code writes the results to a plain text file in tabulated format.

The control file is a user-generated file specifying the EFT interaction, and any optional customizations to the settings. (See~\ref{sec: cfile}.)
The nuclear structure inputs needed are one-body density matrix files (See~\ref{nuclides}). 

An example input file can be found in the runs directory. The file \texttt{example.input} contains:
\begin{verbatim}
er
54
77
example
../targets/Xe/xe131gcn
1.0  1000.0  1.0
\end{verbatim}
These inputs would compute a [er] differential event rate spectra; for [54] Xenon; [77] mass number 131 (54+77); [example] using the control file example.control located in the current directory; [../targets/Xe/xe131gcn] using the density matrix file xe131gcn.dres located in the \texttt{targets/Xe} directory; [1.0 1000.0 1.0] for a range of recoil energies from 1 to 1000 keV in 1 keV increments.

The corresponding control file, \texttt{example.control}, contains:
\begin{verbatim}
wimpmass     150.0
coefnonrel   1  n          0.00048
...
\end{verbatim}
with 29 additional lines setting the remaining coefficients to equivalent values. This combination of input and control file produces the calculation in the last row of Table \ref{tab:timing}.

Assuming a user has successfully compiled the code, they can run this example from the `runs' directory like this:
\begin{verbatim}
../bin/dmscatter < example.input
\end{verbatim}

While the Fortran {\tt dmscatter} executable can be run by itself, we provide Python application programming interfaces (APIs) for integrating the Fortran program into Python work flows. See section \ref{sec: wrappper} for details.

\section{Python interface}\label{sec: wrappper}

We provide a Python interface (a wrapper) for the Fortran code and a number of example scripts demonstrating its use. The wrapper comes with two Python functions EventrateSpectra, and NucFormFactor which can be imported from dmscatter.py in the Python directory. 
Each function has three required arguments: 
\begin{enumerate}
    \item Number of protons Z in the target nucleus
    \item Number of neutrons N in the target nucleus
    \item Nuclear structure file name (.dres)
\end{enumerate}
If no other arguments are provided, default values will be used for all of the remaining necessary parameters, including zero interaction strength. Default values are specified in the control file keyword table (see~\ref{sec: cwords}).

\subsection{Event rate spectra}

To calculate an event rate with a nonzero interaction, the user should also provide one or more of the optional EFT coupling coefficient arrays: \texttt{cp, cn, cs, cv}. These set the couplings to protons, neutrons, isoscalar, and isovector, respectively. The $0^{th}$ index sets the first operator coefficient: \texttt{cp[0]}$= c_1^p$, etc.
Finally, the user can also pass a dictionary of valid control keywords and values to the function in order to set any of the control words defined in~\ref{sec: cwords}.

To compute the event-rate spectra for $^{131}$Xe with a WIMP mass of 50 GeV and a $c_3^v=0.0048$ coupling, one might call:
\begin{lstlisting}[language=Python]
import dmscatter as dm
control_dict = {"wimpmass" : 50.0}
cv = np.zeros(15)
cv[2] = 0.0048 
Erkev, ER = dm.EventrateSpectra(
            Z = 54,
            N = 77,
            dres = "../targets/Xe/xe131gcn",
            controlwords = control_dict,
            cv = cv,
            exec_path = "../bin/dmscatter")
\end{lstlisting}
This will return the differential event rate spectra for recoil energies from 1 keV to 1 MeV in 1 keV steps. 

The file `xe131gcn.dres' must be accessible at the relative or absolute path name specified (in this case `../targets/Xe/'), and contain a valid one-body density matrix for $^{131}$Xe. Similarly, the {\tt dmscatter} executable 
should be accessible from the user's default path - or else the path to the executable should be specified, as in the above example (\texttt{exec\_path =} 
\texttt{ "../bin/dmscatter"}).

\subsection{Nuclear response functions}

We have provided an additional option in {\tt dmscatter} which computes these nuclear form factors from the target density-matrix and exports the results to a data file. This output may be useful for codes like WimPyDD \cite{jeong2021wimpydd}, which compute the WIMP-nucleus event rate spectra starting from nuclear form factors (equation (\ref{responsefunctions})) from an external source. (The DDCalc dark matter direct detection phenomenology package~\cite{bringmann2017darkbit,athron2019global} does have a fast 
Fortran90 central engine and  predicts  signal rates and likelihoods 
for ongoing 
dark matter experiments, but, unlike {\tt dmscatter}, the nuclear structure 
input for the current version of DDCalc is fixed.)

The Python wrapper-function \texttt{NucFormFactor} runs the dmscatter option to export the nuclear response functions to file, and additionally creates and returns an interpolation function $W(q)$ which can be called. In the following code listing, the nuclear response function for $^{131}$Xe is generated for transfer momentum from 0.001 to 10.0 GeV/c. 
\begin{lstlisting}[language=Python]
import dmscatter as dm
cwords = {"usemomentum": 1} 
Wfunc = dm.NucFormFactor(
        Z = 54,
        N = 77,
        dres = "../targets/Xe/xe131gcn",
        controlwords = cwords,
        epmin = 0.001,
        epmax = 10.0,
        epstep = 0.001)
Wfunc(0.001)
\end{lstlisting}
The final line returns an (8,2,2)-shaped array with the evaluate nuclear response functions at $q=0.001$ GeV/c. Note that, had we not set the keyword \texttt{usemomentum} to 1, the function input values would have been specified in terms of recoil energy (the default) instead of transfer momentum.

\section{Performance}

The Fortran code has been optimized for multi-processor CPUs with shared memory architecture using OpenMP. As a result of this parallelism and the inherent efficiency of a compiled language, our program sees an extreme speedup when computing event-rate spectra compared to the Mathematica package \texttt{dmformfactor} version 6. 

We provide timing data for two benchmark cases: $^{29}$Si and $^{131}$Xe, shown in Table \ref{tab:timing}. The compute time of our code depends primarily on two sets of factors: the first is the number of elements in the nuclear densities matrices (which depends on the complexity of the nuclear structure for a given target nucleus), and the second is the number of nonzero EFT coefficients. We include logic to skip compute cycles over zero EFT coefficients.
\begin{table}[h!]
    \centering
    \begin{tabular}{l  c  r  r  r}
        Target &Coupling & {dmformfactor}& 
        \multicolumn{2}{c}{{This work}}\\
        & & & Serial & Parallel\\
        \hline
        \multirow{5}{*}{$^{29}$Si} &$c_1^n$ & 3,800 & 0.5 & 0.2 \\
        &$c_3^n$ & 3,800 & 1.3 & 0.5 \\
        &$c_4^n$ & 3,700 & 1.5 & 0.5 \\
        &$c_5^n$ & 3,700 & 0.9 & 0.3 \\
        &$c_6^n$ & 3,700 & 0.8 & 0.3\\
        \hline
        \multirow{5}{*}{$^{131}$Xe} &$c_1^n$ & 20,000 & 1.7 & 0.5 \\
        &$c_1^p$ &        & 1.7 & 0.5 \\
        &$c_1^s$ & 20,000 & 5.8 & 1.6 \\
        &$c_1^v$ &        & 5.8 & 1.6 \\
        & All & 20,000 & 75 & 20 \\
    \end{tabular}
    \caption{Program execution time in seconds for a sample event-rate calculation with 1000 recoil energies with $m_\chi=$ 150 GeV. The velocity distribution was taken to be Maxwellian with $v_{escape}\approx \infty$. All calculations were done on the same machine (Apple M1 processor, 2020). Multi-threaded execution was performed with 4 threads on the 8-core CPU. $^{29}$Si has 23 matrix elements in its one-body density matrix, while $^{131}$Xe has 67.}
    \label{tab:timing}
\end{table}

\begin{figure}
    \centering
    \includegraphics[width=\columnwidth]{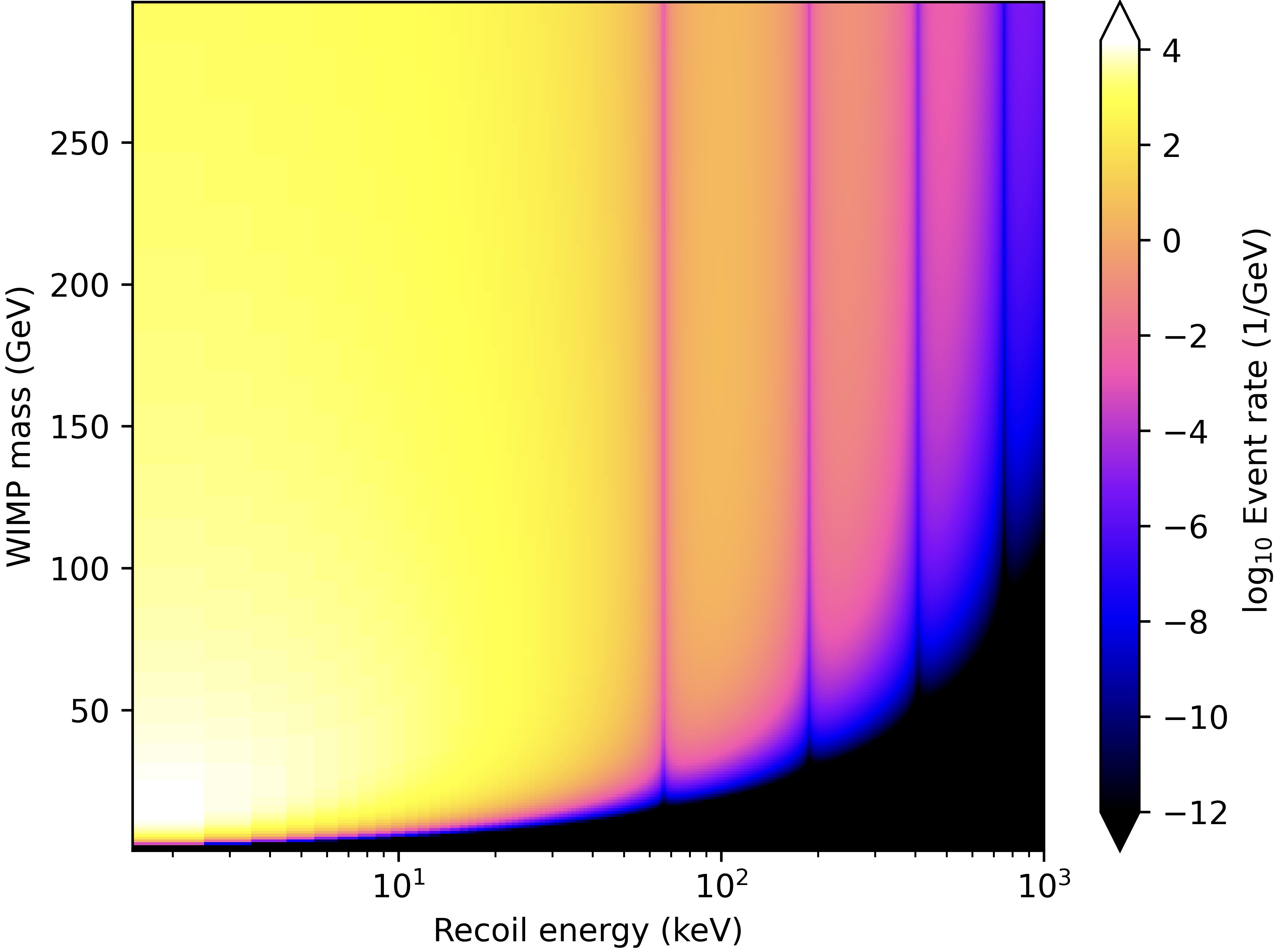}
    \caption{Event rate as a function of WIMP mass and recoil energy for $^{131}$Xe. 300 masses are represented in this figure (1000 recoil energies each; a slice along the horizontal axis is analogous to the calculation shown in Figure \ref{fig: eventrate}). Using our code, this consumed roughly 30 minutes of CPU time. We estimate that using {\tt dmformfactor} to generate the same figure would take at least 70 days (300 curves, 20 000 seconds each). The EFT interaction used was vector isospin-coupling to operator-1 ($c_1^{\tau=1}=0.00048$).}
    \label{fig:wimp_mass}
\end{figure}

The timing data in Table \ref{tab:timing} provides a general indication of the compute time for basic calculations. We also ran a more complex benchmark calculation  to  represent the complexity of a practical application. In this calculation, we compute the differential event rate spectra for $^{131}$Xe over a range of recoil energies from 1 keV to 1000 keV, and for a range of WIMP masses from 1 GeV to 300 GeV, in 1 GeV increments. We provide the Python script \texttt{exampleMassHeatPlot.py} used to generate this plot in the \texttt{python/} directory. The result is shown in the heat plot in Figure \ref{fig:wimp_mass}. This calculation represents 300 calculations of the type in Table \ref{tab:timing}, and so we estimate that generating the data for such a plot using the Mathematica package \texttt{dmformfactor} would require roughly 70 days of CPU time. Our calculation takes only 20 minutes of CPU time with serial execution (including overhead from the Python script running the code), and with parallel execution across 4 threads the wall time is 7 minutes.

\section{Application to experimental limit setting}

Adapting the scripts provided with this code, we generate a bank of energy spectra for a sample of WIMP masses, both nucleons, and all operators. We can then build up a model for a liquid xenon Time Projection Chamber (TPC), following the background model from \cite{Akerib_2020LZ} for both electronic and nuclear recoils (ER and NR respectively) as a function of energy. Since the WIMP interactions are nuclear only, we then need to convert the observed electronic recoils into the equivalent number of events seen in the NR band. To accomplish this, we apply the discrimination as a function of energy reported by the Large Underground Xenon experiment (LUX) \cite{Akerib_2020Discrimination} which has an average discrimination of $\mathcal{O}(10^{-3})$ at energies from 0 to 9.7 keVee. This discrimination factor is slightly optimistic as it assumes perfect corrections. Comparison of ER and NR backgrounds cannot be done directly as the energy scales differ between the two interaction types. To account for this we apply a scaling factor to the nuclear recoils to bring them both into an electron equivalent scale: $E_{ee}=0.173\,E_{nr}^{1.05}$. We limit the search region to 120 keV as above that energy more careful handling of background components would be required.

Once both the background and the signal components can be compared directly, we build up a likelihood function, $\mathcal{L}$.  
\begin{multline}
    \mathcal{L}(\mu_s,{\vec{\mu}}|\mathcal{D}) = \\ \left[ \text{Pois}(n_0|\mu)\prod_{e=1}^{n_0}\frac{1}{\mu_{tot}}\left(\mu_{s}f_{s}(E)+\sum_{i}\mu_if_i(E)\right)\right] \\
    \times \prod_{p}f_p({\bf{g_p}}|\mu_p)
\end{multline}
where the dataset $\mathcal{D}$ contains $n_0$ events, the set of nuisance parameters ${\vec{\mu}}$ represents the number of events from a given component scaling the normalized energy PDF $f_i(E)$.  We also apply a constraint term $f_p({\bf{g_p}}|\mu_p)$ which is a Gaussian term relating the fit value $\mu_p$ to an external constraint ${\bf{g_p}}$. We use Gaussian widths in the constraint term of 20\% on both ER and NR components following \cite{Akerib_2020LZ}.

Using this model, we run a profile likelihood ratio (PLR) based limit setting technique with a two-sided test statistic. Following the prescription of Cowan \cite{Cowan_2011}
we compute the asymptotic limit on an ``Asimov dataset,'' which has all nuisance parameters set to their expectations. For more details on this method, see the statistics section of~\cite{Ohare_2021}, which follows a similar method. The median background-only test statistic can be approximated as the test statistic of the Asimov dataset. Then the p-value at a given number of signal events can be computed from the integral of the asymptotic approximation to the signal plus background test statistic distribution. The 90\% upper limit can then be computed by finding the number of signal events that corresponds to a p-value of 90\%. The final limit can then be computed as the ratio of the 90\% upper limit on the number of signal events divided by the integrated signal flux. 
\begin{figure*}
    \centering
    \includegraphics[width=0.95\textwidth]{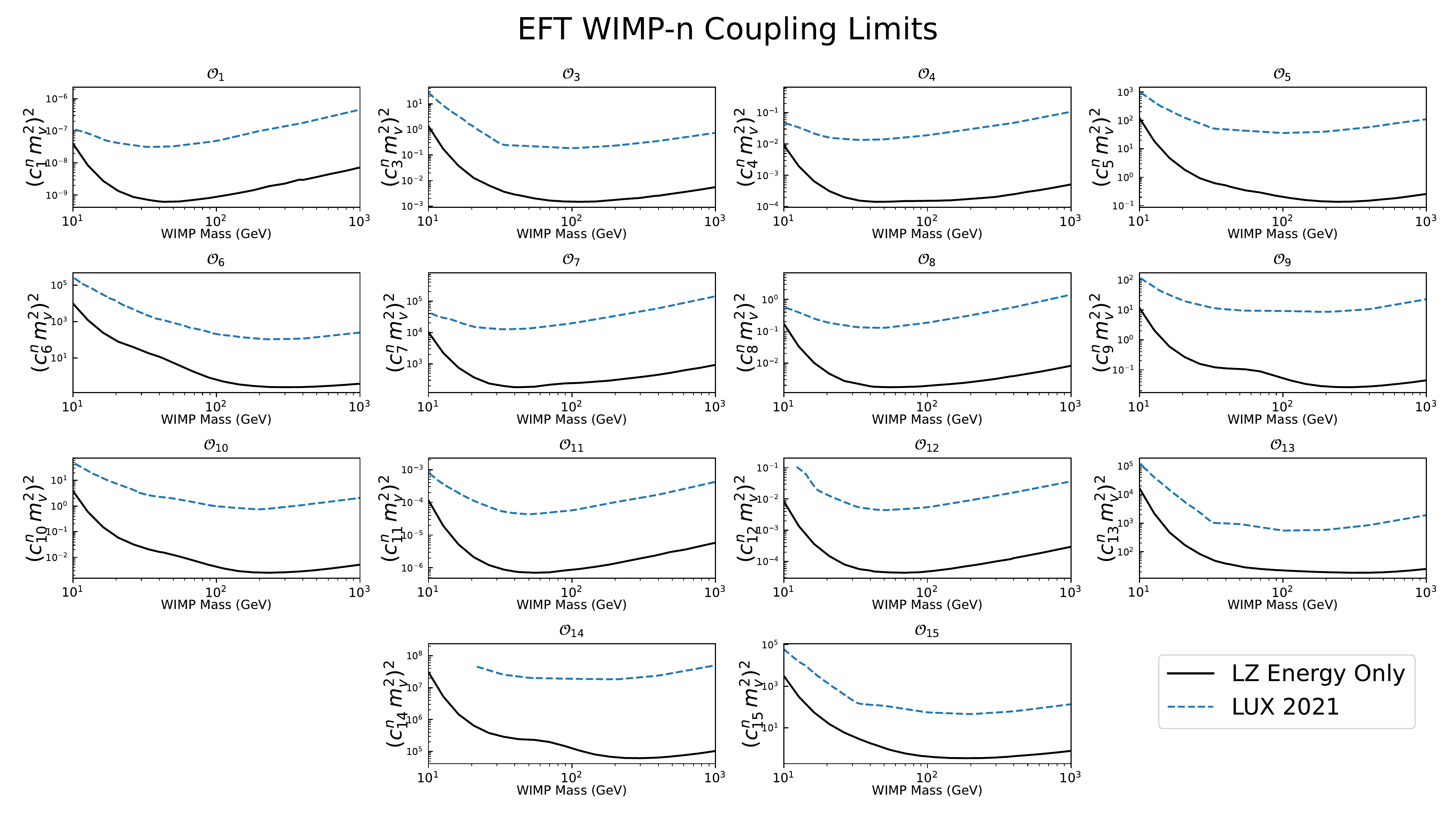}
    \caption{90\% CL limits on the dimensionless WIMP-neutron coupling ($c_i^n$) for each of the 14 non-relativistic EFT operators. Each panel shows the limit produced using the method outlined in the text scaled to the LZ exposure of $5.6\times10^6$ kg$\cdot$day exposure in solid black and the most recent result from the LUX experiment with $3.14\times10^4$ kg$\cdot$day exposure in dashed blue \cite{akerib2020effective}.}
    \label{fig:wimpn}
\end{figure*}
\begin{figure*}
    \centering
    \includegraphics[width=0.95\textwidth]{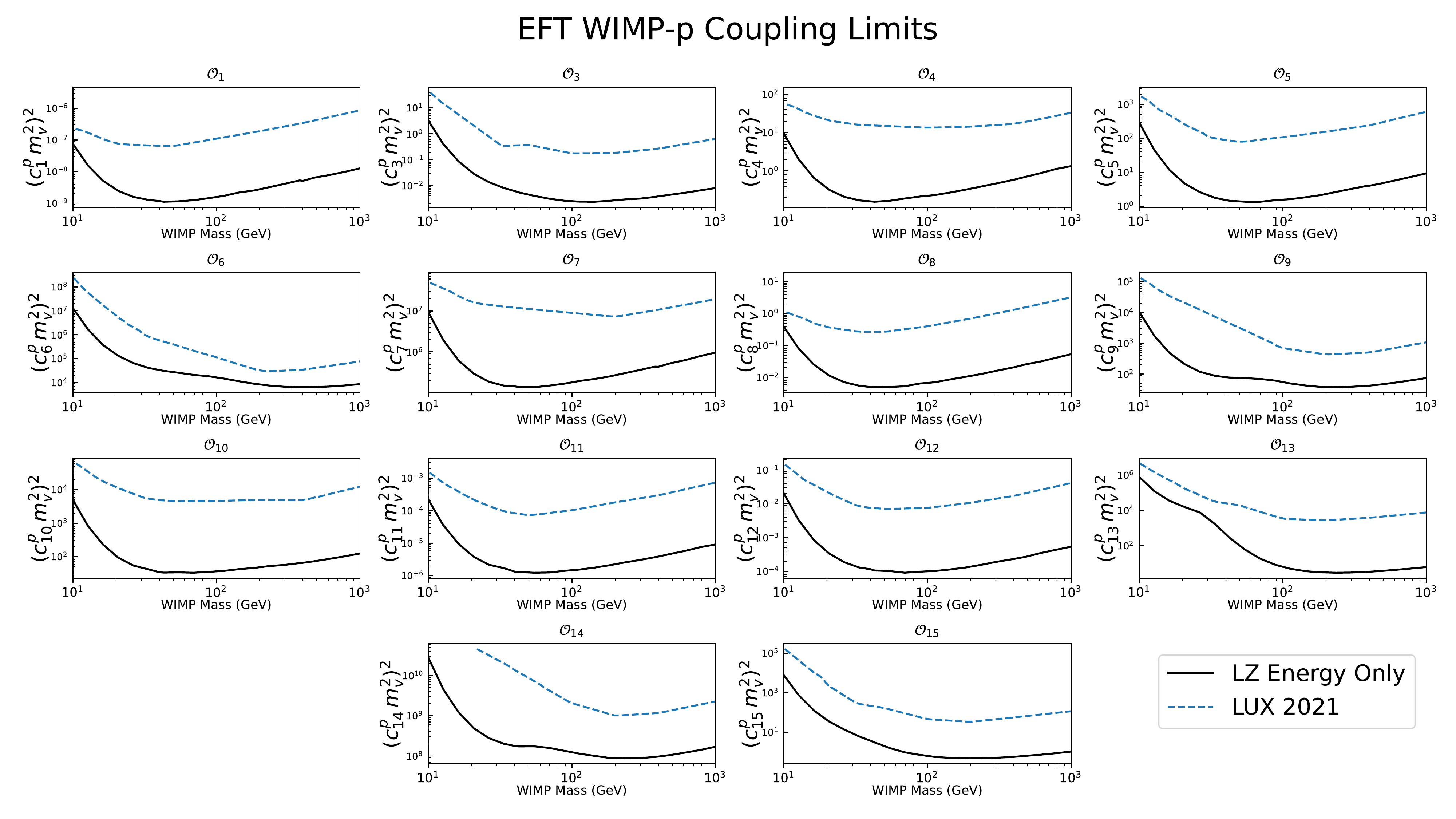}
    \caption{Same as Figure \ref{fig:wimpn} for limits on the dimensionless WIMP-proton coupling ($c_i^p$).}
    \label{fig:wimpp}
\end{figure*}
The results of this asymptotic PLR method are shown in Figures \ref{fig:wimpn} and \ref{fig:wimpp}; note that the coupling here is scaled to the Higgs vacuum expectation value squared, as this is the internal scaling factor of both this code and \cite{Anand_2014}. The relative scaling in these figures is consistent with the scaling due to the increased exposure of the LZ experiment compared to the LUX experiment. The difference at low mass arises because of the conservative choice of the threshold required to do an energy-only search.

\section{Conclusion}

Our modern Fortran code implements an established effective field theory framework for computing WIMP-nucleus form factors, offering performance that is roughly three orders of magnitude faster than the available Mathematica implementation. This speedup, along with the accompanying Python APIs, will enable researchers to explore the EFT parameter space with an efficiency and level of detail that was previously impractical due to the demanding computational cost.

While there exist similar scripts in Python for speeding up such searches~\cite{kang2020sensitivity,jeong2021wimpydd}, they suppose pre-computed nuclear form factors (e.g. as computed by \texttt{dmformfactor}) as inputs and provide a means to test different EFT couplings. Our code computes everything starting from the EFT coefficients and the nuclear wave function in a density matrix format. This means new targets can easily be implemented just as quickly, and sensitivity studies can be performed against the nuclear structure uncertainties.

\section{Acknowledgements}

We gratefully acknowledge the pioneering Mathematica script of Anand, Fitzsimmons, and Haxton, which made the job of validating our code much easier. For useful feedback on the code and paper, we thank  S. Alsum, J. Green, W. C. Haxton, K. Palladino; in particular we thanks A. Suliga, and B. Lem for helping us find bugs and inconsistencies.   We also thank A.~B.~Balantekin and S.~N.~Coppersmith for encouragement and support of this work. 
We thank F.~Kahlhoefer for bring the DDCalc package~\cite{bringmann2017darkbit,athron2019global} to our attention.
Finally, we thank W.~C.~Haxton and K.~S.~McElvain for sharing their extensive library of density matrices for dark matter targets.

This work was supported by the U.S. Department of Energy, Office of Science, Office of High Energy Physics, under Award Number DE-SC0019465. The code is free and open-source, released under the MIT License. 

\bibliography{dmscatter}

\begin{thebibliography}{10}
\expandafter\ifx\csname url\endcsname\relax
  \def\url#1{\texttt{#1}}\fi
\expandafter\ifx\csname urlprefix\endcsname\relax\def\urlprefix{URL }\fi
\expandafter\ifx\csname href\endcsname\relax
  \def\href#1#2{#2} \def\path#1{#1}\fi

\bibitem{Fitzpatrick_2013}
A.~L. Fitzpatrick, W.~Haxton, E.~Katz, N.~Lubbers, Y.~Xu,
  \href{https://doi.org/10.1088/1475-7516/2013/02/004}{The effective field
  theory of dark matter direct detection}, Journal of Cosmology and
  Astroparticle Physics 2013~(02) (2013) 004--004.
\newblock \href {https://doi.org/10.1088/1475-7516/2013/02/004}
  {\path{doi:10.1088/1475-7516/2013/02/004}}.
\newline\urlprefix\url{https://doi.org/10.1088/1475-7516/2013/02/004}

\bibitem{Anand_2014}
N.~Anand, A.~L. Fitzpatrick, W.~C. Haxton,
  \href{http://dx.doi.org/10.1103/PhysRevC.89.065501}{Weakly interacting
  massive particle-nucleus elastic scattering response}, Physical Review C
  89~(6) (Jun 2014).
\newblock \href {https://doi.org/10.1103/physrevc.89.065501}
  {\path{doi:10.1103/physrevc.89.065501}}.
\newline\urlprefix\url{http://dx.doi.org/10.1103/PhysRevC.89.065501}

\bibitem{Blumenthal:1984bp}
G.~R. Blumenthal, S.~Faber, J.~R. Primack, M.~J. Rees, {Formation of Galaxies
  and Large Scale Structure with Cold Dark Matter}, Nature 311 (1984) 517--525.
\newblock \href {https://doi.org/10.1038/311517a0}
  {\path{doi:10.1038/311517a0}}.

\bibitem{Bertone:2016nfn}
G.~Bertone, D.~Hooper, {History of dark matter}, Rev. Mod. Phys. 90~(4) (2018)
  045002.
\newblock \href {http://arxiv.org/abs/1605.04909} {\path{arXiv:1605.04909}},
  \href {https://doi.org/10.1103/RevModPhys.90.045002}
  {\path{doi:10.1103/RevModPhys.90.045002}}.

\bibitem{Primack:1988zm}
J.~R. Primack, D.~Seckel, B.~Sadoulet, {Detection of Cosmic Dark Matter}, Ann.
  Rev. Nucl. Part. Sci. 38 (1988) 751--807.
\newblock \href {https://doi.org/10.1146/annurev.ns.38.120188.003535}
  {\path{doi:10.1146/annurev.ns.38.120188.003535}}.

\bibitem{Feng:2010gw}
J.~L. Feng, {Dark Matter Candidates from Particle Physics and Methods of
  Detection}, Ann. Rev. Astron. Astrophys. 48 (2010) 495--545.
\newblock \href {http://arxiv.org/abs/1003.0904} {\path{arXiv:1003.0904}},
  \href {https://doi.org/10.1146/annurev-astro-082708-101659}
  {\path{doi:10.1146/annurev-astro-082708-101659}}.

\bibitem{PhysRevD.31.3059}
M.~W. Goodman, E.~Witten,
  \href{https://link.aps.org/doi/10.1103/PhysRevD.31.3059}{Detectability of
  certain dark-matter candidates}, Phys. Rev. D 31 (1985) 3059--3063.
\newblock \href {https://doi.org/10.1103/PhysRevD.31.3059}
  {\path{doi:10.1103/PhysRevD.31.3059}}.
\newline\urlprefix\url{https://link.aps.org/doi/10.1103/PhysRevD.31.3059}

\bibitem{RevModPhys.71.S197}
B.~Sadoulet,
  \href{https://link.aps.org/doi/10.1103/RevModPhys.71.S197}{Deciphering the
  nature of dark matter}, Rev. Mod. Phys. 71 (1999) S197--S204.
\newblock \href {https://doi.org/10.1103/RevModPhys.71.S197}
  {\path{doi:10.1103/RevModPhys.71.S197}}.
\newline\urlprefix\url{https://link.aps.org/doi/10.1103/RevModPhys.71.S197}

\bibitem{Vietze:2014vsa}
L.~Vietze, P.~Klos, J.~Men\'endez, W.~Haxton, A.~Schwenk, {Nuclear structure
  aspects of spin-independent WIMP scattering off xenon}, Phys. Rev. D 91~(4)
  (2015) 043520.
\newblock \href {http://arxiv.org/abs/1412.6091} {\path{arXiv:1412.6091}},
  \href {https://doi.org/10.1103/PhysRevD.91.043520}
  {\path{doi:10.1103/PhysRevD.91.043520}}.

\bibitem{Hoferichter:2020osn}
M.~Hoferichter, J.~Men\'endez, A.~Schwenk, {Coherent elastic neutrino-nucleus
  scattering: EFT analysis and nuclear responses}, Phys. Rev. D 102 (2020)
  074018.
\newblock \href {http://arxiv.org/abs/2007.08529} {\path{arXiv:2007.08529}}.

\bibitem{HaxtonNewCode}
W.~C. Haxton, unpublished.

\bibitem{xia2019pandax}
J.~Xia, A.~Abdukerim, W.~Chen, X.~Chen, Y.~Chen, X.~Cui, D.~Fang, C.~Fu,
  K.~Giboni, F.~Giuliani, et~al., Pandax-{II} constraints on spin-dependent
  {WIMP}-nucleon effective interactions, Physics Letters B 792 (2019) 193--198.

\bibitem{DMreview}
W.~C. Haxton, C.~W. Johnson, K.~S. McElvain, to be submitted to Annual Review
  of Nuclear and Particle Science (2022).

\bibitem{PhysRevC.101.054308}
J.~M.~R. Fox, C.~W. Johnson, R.~N. Perez,
  \href{https://link.aps.org/doi/10.1103/PhysRevC.101.054308}{Uncertainty
  quantification of an empirical shell-model interaction using principal
  component analysis}, Phys. Rev. C 101 (2020) 054308.
\newblock \href {https://doi.org/10.1103/PhysRevC.101.054308}
  {\path{doi:10.1103/PhysRevC.101.054308}}.
\newline\urlprefix\url{https://link.aps.org/doi/10.1103/PhysRevC.101.054308}

\bibitem{bringmann2017darkbit}
T.~Bringmann, J.~Conrad, J.~M. Cornell, L.~A. Dal, J.~Edsj{\"o}, B.~Farmer,
  F.~Kahlhoefer, A.~Kvellestad, A.~Putze, C.~Savage, et~al., Darkbit: a gambit
  module for computing dark matter observables and likelihoods, The European
  Physical Journal C 77~(12) (2017) 1--57.

\bibitem{athron2019global}
P.~Athron, C.~Bal{\'a}zs, A.~Beniwal, S.~Bloor, J.~E. Camargo-Molina, J.~M.
  Cornell, B.~Farmer, A.~Fowlie, T.~E~Gonzalo, F.~Kahlhoefer, et~al., Global
  analyses of higgs portal singlet dark matter models using gambit, The
  European Physical Journal C 79~(1) (2019) 1--28.

\bibitem{PhysRevD.33.3495}
A.~K. Drukier, K.~Freese, D.~N. Spergel,
  \href{https://link.aps.org/doi/10.1103/PhysRevD.33.3495}{Detecting cold
  dark-matter candidates}, Phys. Rev. D 33 (1986) 3495--3508.
\newblock \href {https://doi.org/10.1103/PhysRevD.33.3495}
  {\path{doi:10.1103/PhysRevD.33.3495}}.
\newline\urlprefix\url{https://link.aps.org/doi/10.1103/PhysRevD.33.3495}

\bibitem{PhysRevD.37.3388}
K.~Freese, J.~Frieman, A.~Gould,
  \href{https://link.aps.org/doi/10.1103/PhysRevD.37.3388}{Signal modulation in
  cold-dark-matter detection}, Phys. Rev. D 37 (1988) 3388--3405.
\newblock \href {https://doi.org/10.1103/PhysRevD.37.3388}
  {\path{doi:10.1103/PhysRevD.37.3388}}.
\newline\urlprefix\url{https://link.aps.org/doi/10.1103/PhysRevD.37.3388}

\bibitem{evans2018shm}
N.~W. Evans, C.~A.~J. O'Hare, C.~McCabe, Shm$^{++}$: A refinement of the
  standard halo model for dark matter searches in light of the gaia sausage
  (2018).
\newblock \href {http://arxiv.org/abs/1810.11468} {\path{arXiv:1810.11468}}.

\bibitem{edmonds1996angular}
A.~R. Edmonds, Angular momentum in quantum mechanics, Princeton University
  Press, 1996.

\bibitem{DONNELLY1979103}
T.~Donnelly, W.~Haxton,
  \href{https://www.sciencedirect.com/science/article/pii/0092640X79900032}{Multipole
  operators in semileptonic weak and electromagnetic interactions with nuclei:
  Harmonic oscillator single-particle matrix elements}, Atomic Data and Nuclear
  Data Tables 23~(2) (1979) 103--176.
\newblock \href {https://doi.org/https://doi.org/10.1016/0092-640X(79)90003-2}
  {\path{doi:https://doi.org/10.1016/0092-640X(79)90003-2}}.
\newline\urlprefix\url{https://www.sciencedirect.com/science/article/pii/0092640X79900032}

\bibitem{BLOMQVIST1968545}
J.~Blomqvist, A.~Molinari,
  \href{https://www.sciencedirect.com/science/article/pii/0375947468905150}{Collective
  $0-$ vibrations in even spherical nuclei with tensor forces}, Nuclear Physics
  A 106~(3) (1968) 545--569.
\newblock \href {https://doi.org/https://doi.org/10.1016/0375-9474(68)90515-0}
  {\path{doi:https://doi.org/10.1016/0375-9474(68)90515-0}}.
\newline\urlprefix\url{https://www.sciencedirect.com/science/article/pii/0375947468905150}

\bibitem{BIGSTICK1}
C.~W. Johnson, W.~E. Ormand, P.~G. Krastev, Factorization in large-scale
  many-body calculations, Computer Physics Communications 184 (2013)
  2761--2774.

\bibitem{BIGSTICK2}
C.~W. Johnson, W.~E. Ormand, K.~S. McElvain, H.~Shan, Bigstick: A flexible
  configuration-interaction shell-model code (2018).
\newblock \href {http://arxiv.org/abs/1801.08432v1}
  {\path{arXiv:1801.08432v1}}.

\bibitem{PhysRevC.68.041001}
D.~R. Entem, R.~Machleidt,
  \href{https://link.aps.org/doi/10.1103/PhysRevC.68.041001}{Accurate
  charge-dependent nucleon-nucleon potential at fourth order of chiral
  perturbation theory}, Phys. Rev. C 68 (2003) 041001.
\newblock \href {https://doi.org/10.1103/PhysRevC.68.041001}
  {\path{doi:10.1103/PhysRevC.68.041001}}.
\newline\urlprefix\url{https://link.aps.org/doi/10.1103/PhysRevC.68.041001}

\bibitem{shirokov2016n3lo}
A.~Shirokov, I.~Shin, Y.~Kim, M.~Sosonkina, P.~Maris, J.~Vary, N3lo nn
  interaction adjusted to light nuclei in ab exitu approach, Physics Letters B
  761 (2016) 87--91.

\bibitem{cohen1965effective}
S.~Cohen, D.~Kurath, Effective interactions for the 1p shell, Nuclear Physics
  73~(1) (1965) 1--24.

\bibitem{wildenthal1984empirical}
B.~Wildenthal, Empirical strengths of spin operators in nuclei, Progress in
  particle and nuclear physics 11 (1984) 5--51.

\bibitem{brown1988status}
B.~A. Brown, B.~Wildenthal, Status of the nuclear shell model, Annual Review of
  Nuclear and Particle Science 38~(1) (1988) 29--66.

\bibitem{PhysRevC.74.034315}
B.~A. Brown, W.~A. Richter,
  \href{https://link.aps.org/doi/10.1103/PhysRevC.74.034315}{New ``{USD}''
  hamiltonians for the $\mathit{sd}$ shell}, Phys. Rev. C 74 (2006) 034315.
\newblock \href {https://doi.org/10.1103/PhysRevC.74.034315}
  {\path{doi:10.1103/PhysRevC.74.034315}}.
\newline\urlprefix\url{https://link.aps.org/doi/10.1103/PhysRevC.74.034315}

\bibitem{PhysRevC.86.051301}
Y.~Utsuno, T.~Otsuka, B.~A. Brown, M.~Honma, T.~Mizusaki, N.~Shimizu,
  \href{https://link.aps.org/doi/10.1103/PhysRevC.86.051301}{Shape transitions
  in exotic si and s isotopes and tensor-force-driven {J}ahn-{T}eller effect},
  Phys. Rev. C 86 (2012) 051301.
\newblock \href {https://doi.org/10.1103/PhysRevC.86.051301}
  {\path{doi:10.1103/PhysRevC.86.051301}}.
\newline\urlprefix\url{https://link.aps.org/doi/10.1103/PhysRevC.86.051301}

\bibitem{PhysRevC.80.064323}
M.~Honma, T.~Otsuka, T.~Mizusaki, M.~Hjorth-Jensen,
  \href{https://link.aps.org/doi/10.1103/PhysRevC.80.064323}{New effective
  interaction for ${f}_{5}{\mathit{pg}}_{9}$-shell nuclei}, Phys. Rev. C 80
  (2009) 064323.
\newblock \href {https://doi.org/10.1103/PhysRevC.80.064323}
  {\path{doi:10.1103/PhysRevC.80.064323}}.
\newline\urlprefix\url{https://link.aps.org/doi/10.1103/PhysRevC.80.064323}

\bibitem{PhysRevC.71.044317}
B.~A. Brown, N.~J. Stone, J.~R. Stone, I.~S. Towner, M.~Hjorth-Jensen,
  \href{https://link.aps.org/doi/10.1103/PhysRevC.71.044317}{Magnetic moments
  of the ${2}_{1}^{+}$ states around $^{132}\mathrm{Sn}$}, Phys. Rev. C 71
  (2005) 044317.
\newblock \href {https://doi.org/10.1103/PhysRevC.71.044317}
  {\path{doi:10.1103/PhysRevC.71.044317}}.
\newline\urlprefix\url{https://link.aps.org/doi/10.1103/PhysRevC.71.044317}

\bibitem{GCN5082}
A.~Gniady, E.~Caurier, F.~Nowacki, Unpublished.

\bibitem{PhysRevLett.100.052503}
E.~Caurier, J.~Men\'endez, F.~Nowacki, A.~Poves,
  \href{https://link.aps.org/doi/10.1103/PhysRevLett.100.052503}{Influence of
  pairing on the nuclear matrix elements of the neutrinoless
  $\ensuremath{\beta}\ensuremath{\beta}$ decays}, Phys. Rev. Lett. 100 (2008)
  052503.
\newblock \href {https://doi.org/10.1103/PhysRevLett.100.052503}
  {\path{doi:10.1103/PhysRevLett.100.052503}}.
\newline\urlprefix\url{https://link.aps.org/doi/10.1103/PhysRevLett.100.052503}

\bibitem{PhysRevC.82.064304}
E.~Caurier, F.~Nowacki, A.~Poves, K.~Sieja,
  \href{https://link.aps.org/doi/10.1103/PhysRevC.82.064304}{Collectivity in
  the light xenon isotopes: A shell model study}, Phys. Rev. C 82 (2010)
  064304.
\newblock \href {https://doi.org/10.1103/PhysRevC.82.064304}
  {\path{doi:10.1103/PhysRevC.82.064304}}.
\newline\urlprefix\url{https://link.aps.org/doi/10.1103/PhysRevC.82.064304}

\bibitem{jeong2021wimpydd}
I.~Jeong, S.~Kang, S.~Scopel, G.~Tomar, Wimpydd: an object-oriented python code
  for the calculation of wimp direct detection signals, arXiv preprint
  arXiv:2106.06207 (2021).

\bibitem{Akerib_2020LZ}
D.~Akerib, C.~Akerlof, S.~Alsum, H.~Araújo, M.~Arthurs, X.~Bai, A.~Bailey,
  J.~Balajthy, S.~Balashov, D.~Bauer, et~al.,
  \href{http://dx.doi.org/10.1103/PhysRevD.101.052002}{Projected wimp
  sensitivity of the lux-zeplin dark matter experiment}, Physical Review D
  101~(5) (Mar 2020).
\newblock \href {https://doi.org/10.1103/physrevd.101.052002}
  {\path{doi:10.1103/physrevd.101.052002}}.
\newline\urlprefix\url{http://dx.doi.org/10.1103/PhysRevD.101.052002}

\bibitem{Akerib_2020Discrimination}
D.~Akerib, S.~Alsum, H.~Araújo, X.~Bai, J.~Balajthy, A.~Baxter, E.~Bernard,
  A.~Bernstein, T.~Biesiadzinski, E.~Boulton, et~al.,
  \href{http://dx.doi.org/10.1103/PhysRevD.102.112002}{Discrimination of
  electronic recoils from nuclear recoils in two-phase xenon time projection
  chambers}, Physical Review D 102~(11) (Dec 2020).
\newblock \href {https://doi.org/10.1103/physrevd.102.112002}
  {\path{doi:10.1103/physrevd.102.112002}}.
\newline\urlprefix\url{http://dx.doi.org/10.1103/PhysRevD.102.112002}

\bibitem{Cowan_2011}
G.~Cowan, K.~Cranmer, E.~Gross, O.~Vitells,
  \href{http://dx.doi.org/10.1140/epjc/s10052-011-1554-0}{Asymptotic formulae
  for likelihood-based tests of new physics}, The European Physical Journal C
  71~(2) (Feb 2011).
\newblock \href {https://doi.org/10.1140/epjc/s10052-011-1554-0}
  {\path{doi:10.1140/epjc/s10052-011-1554-0}}.
\newline\urlprefix\url{http://dx.doi.org/10.1140/epjc/s10052-011-1554-0}

\bibitem{Ohare_2021}
C.~A.~J. O’Hare, \href{http://dx.doi.org/10.1103/PhysRevLett.127.251802}{New
  definition of the neutrino floor for direct dark matter searches}, Physical
  Review Letters 127~(25) (Dec 2021).
\newblock \href {https://doi.org/10.1103/physrevlett.127.251802}
  {\path{doi:10.1103/physrevlett.127.251802}}.
\newline\urlprefix\url{http://dx.doi.org/10.1103/PhysRevLett.127.251802}

\bibitem{akerib2020effective}
D.~S. Akerib, S.~Alsum, H.~M. Araújo, X.~Bai, J.~Balajthy, A.~Baxter, E.~P.
  Bernard, A.~Bernstein, T.~P. Biesiadzinski, E.~M. Boulton, B.~Boxer,
  P.~Brás, S.~Burdin, D.~Byram, M.~C. Carmona-Benitez, C.~Chan, J.~E. Cutter,
  L.~de~Viveiros, E.~Druszkiewicz, A.~Fan, S.~Fiorucci, R.~J. Gaitskell,
  C.~Ghag, M.~G.~D. Gilchriese, C.~Gwilliam, C.~R. Hall, S.~J. Haselschwardt,
  S.~A. Hertel, D.~P. Hogan, M.~Horn, D.~Q. Huang, C.~M. Ignarra, R.~G.
  Jacobsen, O.~Jahangir, W.~Ji, K.~Kamdin, K.~Kazkaz, D.~Khaitan, E.~V.
  Korolkova, S.~Kravitz, V.~A. Kudryavtsev, N.~A. Larsen, E.~Leason, B.~G.
  Lenardo, K.~T. Lesko, J.~Liao, J.~Lin, A.~Lindote, M.~I. Lopes,
  A.~Manalaysay, R.~L. Mannino, N.~Marangou, D.~N. McKinsey, D.~M. Mei,
  M.~Moongweluwan, J.~A. Morad, A.~S.~J. Murphy, A.~Naylor, C.~Nehrkorn, H.~N.
  Nelson, F.~Neves, A.~Nilima, K.~C. Oliver-Mallory, K.~J. Palladino, E.~K.
  Pease, Q.~Riffard, G.~R.~C. Rischbieter, C.~Rhyne, P.~Rossiter, S.~Shaw,
  T.~A. Shutt, C.~Silva, M.~Solmaz, V.~N. Solovov, P.~Sorensen, T.~J. Sumner,
  M.~Szydagis, D.~J. Taylor, R.~Taylor, W.~C. Taylor, B.~P. Tennyson, P.~A.
  Terman, D.~R. Tiedt, W.~H. To, L.~Tvrznikova, U.~Utku, S.~Uvarov,
  A.~Vacheret, V.~Velan, R.~C. Webb, J.~T. White, T.~J. Whitis, M.~S.
  Witherell, F.~L.~H. Wolfs, D.~Woodward, J.~Xu, C.~Zhang, An effective field
  theory analysis of the first lux dark matter search (2020).
\newblock \href {http://arxiv.org/abs/2003.11141} {\path{arXiv:2003.11141}}.

\bibitem{kang2020sensitivity}
S.~Kang, S.~Scopel, G.~Tomar, J.-H. Yoon, On the sensitivity of present direct
  detection experiments to wimp--quark and wimp--gluon effective interactions:
  A systematic assessment and new model--independent approaches, Astroparticle
  Physics 114 (2020) 80--91.

\bibitem{LEWIN199687}
J.~Lewin, P.~Smith,
  \href{https://www.sciencedirect.com/science/article/pii/S0927650596000473}{Review
  of mathematics, numerical factors, and corrections for dark matter
  experiments based on elastic nuclear recoil}, Astroparticle Physics 6~(1)
  (1996) 87--112.
\newblock \href {https://doi.org/https://doi.org/10.1016/S0927-6505(96)00047-3}
  {\path{doi:https://doi.org/10.1016/S0927-6505(96)00047-3}}.
\newline\urlprefix\url{https://www.sciencedirect.com/science/article/pii/S0927650596000473}

\bibitem{PhysRevD.82.075004}
A.~L. Fitzpatrick, K.~M. Zurek,
  \href{https://link.aps.org/doi/10.1103/PhysRevD.82.075004}{Dark moments and
  the dama-cogent puzzle}, Phys. Rev. D 82 (2010) 075004.
\newblock \href {https://doi.org/10.1103/PhysRevD.82.075004}
  {\path{doi:10.1103/PhysRevD.82.075004}}.
\newline\urlprefix\url{https://link.aps.org/doi/10.1103/PhysRevD.82.075004}

\bibitem{PhysRevD.82.023530}
C.~McCabe,
  \href{https://link.aps.org/doi/10.1103/PhysRevD.82.023530}{Astrophysical
  uncertainties of dark matter direct detection experiments}, Phys. Rev. D 82
  (2010) 023530.
\newblock \href {https://doi.org/10.1103/PhysRevD.82.023530}
  {\path{doi:10.1103/PhysRevD.82.023530}}.
\newline\urlprefix\url{https://link.aps.org/doi/10.1103/PhysRevD.82.023530}

\bibitem{Davis1984}
\href{https://www.sciencedirect.com/science/article/pii/B9780122063602500145}{Appendix
  2 - fortran programs}, in: P.~J. Davis, P.~Rabinowitz (Eds.), Methods of
  Numerical Integration (Second Edition), second edition Edition, Academic
  Press, 1984, pp. 480--508.
\newblock \href
  {https://doi.org/https://doi.org/10.1016/B978-0-12-206360-2.50014-5}
  {\path{doi:https://doi.org/10.1016/B978-0-12-206360-2.50014-5}}.
\newline\urlprefix\url{https://www.sciencedirect.com/science/article/pii/B9780122063602500145}

\bibitem{zhang1996computation}
S.~Zhang, J.~Jin,
  \href{https://books.google.com/books?id=ASfvAAAAMAAJ}{Computation of Special
  Functions}, Wiley, 1996.
\newline\urlprefix\url{https://books.google.com/books?id=ASfvAAAAMAAJ}

\bibitem{PhysRevC.78.064302}
W.~A. Richter, S.~Mkhize, B.~A. Brown,
  \href{https://link.aps.org/doi/10.1103/PhysRevC.78.064302}{$\mathit{sd}$-shell
  observables for the usda and usdb hamiltonians}, Phys. Rev. C 78 (2008)
  064302.
\newblock \href {https://doi.org/10.1103/PhysRevC.78.064302}
  {\path{doi:10.1103/PhysRevC.78.064302}}.
\newline\urlprefix\url{https://link.aps.org/doi/10.1103/PhysRevC.78.064302}

\bibitem{gysbers2019discrepancy}
P.~Gysbers, G.~Hagen, J.~Holt, G.~R. Jansen, T.~D. Morris, P.~Navr{\'a}til,
  T.~Papenbrock, S.~Quaglioni, A.~Schwenk, S.~Stroberg, et~al., Discrepancy
  between experimental and theoretical $\beta$-decay rates resolved from first
  principles, Nature Physics 15~(5) (2019) 428--431.

\end{thebibliography}

\appendix

\section{Control file (.control)}\label{sec: cfile}

Each EFT parameter is written on its own line in `[mycontrolfile].control', with four values: the keyword `coefnonrel', the operator number (integer 1..16), the coupling type (`p'=proton, `n'=neutron, `s'=scalar, `v'=vector), and the coefficient value. For example: 
\begin{verbatim}
coefnonrel    1    s     3.1
\end{verbatim}
would set $c_1^{\tau=0} = 3.1$. We take the isospin convention:
\begin{equation}
	\begin{split}
		c^0 = \frac{1}{2} (c^p + c^n) \\
		c^1 =\frac{1}{2}( c^p - c^n)
	\end{split}
\end{equation}
Thus, the previous example is equivalent to:
\begin{verbatim}
coefnonrel    1    p     1.55
coefnonrel    1    n     1.55
\end{verbatim}

The control file also serves a more general but optional function: to set any parameter in the program to a custom value. Simply add an entry to the control file with two values: the first should be the keyword for the parameter and the second should be the value to set that parameter to. For example, to set the velocity of the earth in the galactic frame to $240\ \mathrm{km/s}$, you should add the line:
\begin{verbatim}
vearth  240.0
\end{verbatim}

As an example, here is the complete control file used to calculate the event rate for the $c^n_1$ coupling to $^{131}$Xe shown in Table \ref{tab:timing}:
\begin{verbatim}
# Coefficient matrix (non-relativistic)
# Ommitted values are assumed to be 0.0.
# c_i^t
# i = 1,...,16
# t: p=proton n=neutron s=scaler v=vector
coefnonrel  1  n  0.00048
wimpmass 150.0
vearth 232.0
maxwellv0 220.0
dmdens 0.3
usemomentum 0
useenergyfile 0
ntscale  2500.0
printdensities 0
#vescape 550.
\end{verbatim}
Un-commenting the last line would set the escape velocity to 550 km/s.

A complete list of control words is given in the manual and in~\ref{sec: cwords}.

\section{Nuclear structure input (.dres)}\label{nuclides}

Users must provide nuclear one-body density matrix elements, either in isospin format,
\begin{equation}
    \rho_{J,T}^{\Psi}(a,b) = (2J+1)^{-1/2}(2T+1)^{-1/2}  \langle \Psi||| [\hat{c}_a^\dagger \otimes \tilde{c}_b]_{J,T}|||\Psi \rangle ,
\end{equation}
or proton-neutron format, Users must provide nuclear one-body density matrix elements, either in isospin form,
\begin{equation}
    \rho_{J}^{\Psi}(a,b) = (2J+1)^{-1/2}  \langle \Psi|| [\hat{c}_a^\dagger \otimes \tilde{c}_b]_{J}||\Psi \rangle ,
\end{equation}
where $\Psi$ is the nuclear-target wave function and $\hat{c}^\dagger$, $\hat{c}$ are the one-body creation, destruction operators. For proton-neutron format, the orbital indices $a$ are distinct for protons and neutrons.  The matrix elements must be stored in a file in a standard format produced by shell-model codes like {\sc Bigstick}. The only assumption is that the single-particle basis states are harmonic oscillator states. If density matrices are generated in some other single-particle basis, such as those from a Woods-Saxon potential or a Hartree-Fock calculation, that basis must be expanded into harmonic oscillator states.  By using harmonic oscillator basis states one can efficiently compute the matrix elements.  One can use either phenomenological or \textit{ab initio} model spaces and interactions; as an example, we provide density matrices for $^{12}$C, both from the phenomenological Cohen-Kurath shell model interaction~\cite{cohen1965effective}, and from two no-core shell model interactions~\cite{PhysRevC.68.041001,shirokov2016n3lo}.  A detailed description of the provenance of the supplied targets can be found in the included manual, and will be updated as more density matrices become available.  We also include, for purposes of validation, the `legacy' density matrices available in the original \texttt{dmformfactor} script.

\subsection{Density matrix format}

We adopt the output format from the {\sc Bigstick} shell-model code. The output one-body densities (\ref{eqn:denmat}) are written to a file with extension \texttt{.dres}. We provide a full specification of this plain-text-file format in the \texttt{docs} directory. Here, we show the form of the file and explain its contents.
\begin{verbatim}
<File header>
 4 -5
  State      E        Ex         J       T
    1   -330.17116   0.00000     1.500  11.500
  Single particle state quantum numbers
ORBIT      N     L   2 x J
     1     0     2     3
     2     0     2     5
     3     1     0     1
 Initial state # 1 E = -330.17117 2xJ, 2xT = 3  23
 Final state   # 1 E = -330.17117 2xJ, 2xT = 3  23
 Jt =   0, proton      neutron
    1    1   1.55844   5.40558
\end{verbatim}
The first line of the file is an arbitrary header. The second line of the file contains two integers: the number of protons and number of neutrons in the valence space. These numbers may be negative, in which case they represent the number of holes in a completely full valence space. In the previous example, there are 4 valence protons and 5 valence neutron holes, which translates to 27 valence neutrons.

The file is thereafter comprised of three sections: 
\begin{enumerate}
    \item many-body state information
    \item single-particle state quantum numbers
    \item density matrix element blocks
\end{enumerate}
Only the ground state is needed for inelastic WIMP-nucleus scattering calculations. The single-particle state quantum numbers specify the quantum numbers for the simple-harmonic oscillator states involved in the one-body operators. Finally, the one-body density matrix elements are listed in nested blocks with three layers:
\begin{itemize}
    \item [i.] the initial and final state specification (corresponding to the many-body states listed in section (1) of the file)
    \item [ii.] the angular momentum carried by the one-body density matrix operator, labeled \texttt{Jt} here
    \item [iii.] the single-particle state labels \texttt{a}, \texttt{b} in columns 1 and 2 (corresponding to the single-particle state labels listed in section (2) of the file) and the proton and neutron (isospin-0 and isospin-1) density matrix elements in columns 3 and 4
\end{itemize}
Both (i) and (ii) must be specified along with columns 1 and 2 of (iii) in order to fully determine a matrix element $\rho^{f,i}_K(a,b)$, where $K=J_t$. Note that the values of $K$ are restricted by conservation of angular momentum; both between the many-body states labeled $i$ and $f$, and the single-particle states labeled $a$ and $b$.

\subsection{Filling core orbitals for phenomenological interactions}

Since standard one-body density matrices in phenomenological model spaces contain only matrix elements for orbitals in the valence space, it is necessary to infer the matrix elements for the core orbitals. Our code does this by default, but the user can disable this option using the \texttt{fillnuclearcore} control word.

For phenomenological interactions one typically has a `frozen' core of nucleons which do not participate in the two-body forces of the Hamiltonian. In such cases the single-particle space listed in the .dres file consists only of the valence orbitals and the one-body density matrices are only specified for the valence orbitals.

{\tt dmscatter} reads the valence space orbitals from the .dres file and infers the number of core nucleons by subtracting the number of valence protons and neutrons from the number of nucleons in the target nucleus. The core orbitals are assumed to be one of the standard shell model orbital sets associated with possible cores: $^4$He, $^{16}$O, $^{40}$Ca,$^{56}$Ni, $^{100}$Sn. 

The one-body density matrix elements for the core orbitals are then determined from the (full) occupation of the core orbitals. Because the orbitals are full and here we assume both proton and neutron orbitals filled, the core can contribute only to $J=0, T=0$ densities. Two formats are possible: proton-neutron format:
\begin{equation}
    \rho_{J,x=p,n}^{\Psi}(a,b)_{(core)} = \delta_{a,b}\delta_{J,0} [j_a],
\end{equation}
where $[y] \equiv \sqrt{2y+1}$ and $j_a$ is the angular momentum of $a$-orbit. $J$ is the total spin of the nuclear target state $\Psi$; and isospin format for a target state with good total isospin $T$:
\begin{align}
    \rho_{J,T=0}^{\Psi}(a,b)_{(core)} &= \delta_{a,b} \delta_{J,0} \delta_{T,0} [1/2][j_a], \\
    \rho_{J,T=1}^{\Psi}(a,b)_{(core)} &= 0.0.
\end{align}
Note that our libraries only include isoscalar cores; if one had cores with $N > Z$ then one could also have $T=1$ contributions. 

\subsection{Calling the library of targets}

To call a nuclide from the library of targets, a mandatory command line input is the name of the \texttt{.dres} file with the nuclear structure information. See Section \ref{runrunrun} above. The location is relative to the \texttt{runs/} directory. Our library of targets is located in directory \texttt{targets/}. To use your own nuclear data file, place it in the same directory or specify the directory relative to the \texttt{runs/} directory, i.e., \texttt{../MyTargets/X} will use \texttt{X.dres} in the directory \texttt{MyTargets/}, where both \texttt{runs/} and \texttt{MyTargets/} are subdirectories of the main directory, \texttt{darkmatter/}.

\section{Theory and Implementation Details}

The theoretical formalism for computing the WIMP-nucleus form factors is largely the same as in \cite{Fitzpatrick_2013,Anand_2014}. In this paper we will therefore only provide the basic formalism which is necessary to understand the differences in our numerical and algorithmic approaches to the implementation.

\subsection{Effective field theory}\label{sec: EFT}

The WIMP-nucleus interaction is defined by the user in terms of an effective field theory Lagrangian, specified implicitly by fifteen operator coupling constants $c_i^x$, (for $i=1,...,15$), where $x=p, n$ for coupling to protons or neutrons individually. The code uses the EFT coefficients in explicit proton-neutron couplings, i.e. the interaction is defined by:
\begin{equation}
    \mathcal{H} = \sum_{x=p,n}\sum_{i=1,15} c^x_i \mathcal{O} _i
\end{equation}
and the 15 momentum-dependent operators are:
\allowdisplaybreaks
\begin{align}
    \mathcal{O} _1 &= 1_\chi 1_N \\
    \mathcal{O} _2 &= (v^\perp)^2  \\
    \mathcal{O} _3 &= i\vec{S}_N \cdot \left(\frac{\vec{q}}{m_N}\times
        \vec{v}^\perp\right)\\
    \mathcal{O} _4 &= \vec{S}_\chi \cdot \vec{S}_N \\
    \mathcal{O} _5 &= i\vec{S}_\chi \cdot \left(\frac{\vec{q}}{m_N}\times
        \vec{v}^\perp\right) \\
    \mathcal{O} _6 &= \left(\vec{S}_\chi \cdot \frac{\vec{q}}{m_N} \right)
        \left(\vec{S}_N \cdot \frac{\vec{q}}{m_N} \right) \\
    \mathcal{O} _7 &= \vec{S}_N\cdot \vec{v}^\perp \\
    \mathcal{O} _8 &= \vec{S}_\chi\cdot \vec{v}^\perp \\
    \mathcal{O} _9 &= i\vec{S}_\chi \cdot \left(\vec{S}_N \times
        \frac{\vec{q}}{m_N}\right)\\
    \mathcal{O} _{10} &= i\vec{S}_N \cdot \frac{\vec{q}}{m_N}\\
    \mathcal{O} _{11} &= i\vec{S}_\chi \cdot \frac{\vec{q}}{m_N}\\
    \mathcal{O} _{12} &= \vec{S}_\chi \cdot \left( \vec{S}_N\times 
        \vec{v}^\perp\right)\\
    \mathcal{O} _{13} &= i\left( \vec{S}_\chi \cdot \vec{v}^\perp \right) 
        \left(\vec{S}_N\cdot \frac{\vec{q}}{m_N}\right )\\
    \mathcal{O} _{14} &= i\left( \vec{S}_\chi \cdot \frac{\vec{q}}{m_N} \right) 
        \left(\vec{S}_N\cdot \vec{v}^\perp \right )\\
    \mathcal{O} _{15} &= -\left(\vec{S}_\chi \cdot \frac{\vec{q}}{m_N} \right )
        \left( \left( \vec{S}_N\times \vec{v}^\perp\right)\cdot 
        \frac{\vec{q}}{m_N} \right)
\end{align}
Operator 2 is generally discarded because it is not a leading order non-relativistic reduction of a manifestly relativistic operator \cite{Anand_2014}.  Operators 1 and 4 correspond to the naive density- and spin-coupling, respectively.  

\subsection{WIMP response functions}\label{sec: wimpresponse}

In the following, the EFT coefficients $c_i^x$ are grouped according to how they couple to each of the eight nuclear responses $W_i^{x,x'}$. As a shorthand, $cl(j) \equiv 4j(j+1)/3$, and $v^{\perp 2}\equiv v^2 - (q/2\mu_t)^2$.
\begin{multline}
   R_{M}^{xx'}(v,q) = \frac{1}{4}cl(j_\chi) [ v^{\perp 2} (c_5^{x}c_5^{x'}q^2 + c_8^{x}c_8^{x'}) + c_{11}^{x}c_{11}^{x'}q^2 ]\\
    + (c_1^{x} + c_2^{x}v^{\perp 2} ) (c_1^{x'} + c_2^{x'}v^{\perp 2} )  
\end{multline}
\begin{multline}             
R_{\Sigma''}^{xx'}(v,q) = \frac{1}{16}cl(j_\chi) [c_6^{x}c_6^{x'}q^4 + (c_{13}^{x}c_{13}^{x'}q^2 + c_{12}^{x} c_{12}^{x'} ) v^{\perp 2}\\
    + 2c_4^xc_6^{x'}q^2 + c_4^xc_4^{x'}] + \frac{1}{4}c_{10}^xc_{10}^{x '}q^2
\end{multline}    
\begin{multline}    
R_{\Sigma'}^{xx'}(v,q) = \frac{1}{32} cl(j_\chi) [ 2c_{9}^{x}c_{9}^{x'}q^2 + ( c_{15}^{x}c_{15}^{x'}q^4 + c_{14}^{x}c_{14}^{x'}q^2 \\
    - 2c_{12}^{x}c_{15}^{x'} q^2 + c_{12}^{x}c_{12}^{x'}) v^{\perp 2} + 2c_{4}^{x}c_{4}^{x'} ]\\
    +\frac{1}{8}(c_{3}^{x}c_3^{x'}q^2 + c_{7}^{x}c_{7}^{x'})v^{\perp 2}
\end{multline}        
\begin{multline}        
R_{\Phi''}^{xx'}(v,q) = \frac{q^2}{16m_N^2}cl(j_\chi) (c_{12}^x - c_{15}^{x}q^2)(c_{12}^{x '}-c_{15}^{x '}q^2 ) \\
    + \frac{q^4}{4m_N^2}c_3^x c_3^{x'} 
\end{multline}    
\begin{multline}    
R_{\tilde{\Phi}'}^{xx'}(v,q) = \frac{q^2}{16m_N^2}cl(j_\chi)(c_{13}^xc_{13}^{x'}q^2 + c_{12}^x c_{12}^{x'})
\end{multline}        
\begin{multline}        
R_{\Delta}^{xx'}(v,q) = \frac{q^2}{4m_N^2}cl(j_\chi) (c_{5}^{x}c_{5}^{x'}q^2 + c_{8}^{x}c_{8}^{x'}) \\
    + 2\frac{q^2}{m_N^2}c_{2}^{x}c_{2}^{x'}v^{\perp 2}
\end{multline}        
\begin{multline}        
R_{\Sigma' \Delta}^{xx'}(v,q) = \frac{q^2}{4m_N}cl(j_\chi) (c_{4}^{x}c_{5}^{x'} - c_{8}^{x}c_{9}^{x'}) - \frac{q^2}{m_N} c_{2}^{x}c_{3}^{x'} v^{\perp 2}
\end{multline}        
\begin{multline}        
R_{M\Phi''}^{xx'}(v,q) = \frac{q^2}{4m_N}cl(j_\chi)c_{11}^{x} (c_{12}^{x'} - c_{15}^{x'} q^2) \\
    + \frac{q^2}{m_N}c_{3}^{x'}  (c_{1}^{x} + c_{2}^{x} v^{\perp 2})
\end{multline}        
It should be noted that the last two dark matter responses are composed entirely of interference terms, which is to say, they do not come into play unless certain combinations  of EFT coefficients are simultaneously active. These are the coefficient pairs listed in Section \ref{validate}. For example, $c_4$ and $c_5$ together will activate $R_{\Sigma' \Delta}$, but not alone.

\subsection{Integrals}\label{sec: integral}

We use numerical quadrature evaluate the integral in equation (\ref{ER}) for the velocity distribution (\ref{eq: f}). While there are analytic solutions \cite{LEWIN199687, PhysRevD.82.075004, PhysRevD.82.023530} for specific forms for the cross section, namely with $v^0$ and $v^2$ dependence, we derive a general equation that can be used for any isotropic dark matter halo model, which to our knowledge has not been presented in publication. The integral has the form:
\begin{equation} \label{I3}
\begin{split}
	I = \int_{\Omega} d^3v \frac{d\sigma(v,q)}{d{q}^2}\ v\ f(\vec{v}+\vec{v}_E),
\end{split}
\end{equation}
where the constraint $\Omega$ is that $v_{min}^2<(\vec{v}+\vec{v}_E)^2<v_{esc}^2$ and $f(\vec{v})$ is equation (\ref{eq: f}). Here we present only the result; the full derivation can be found in the manual. Switching to spherical coordinates and taking special care for the constraint $\Omega$, one obtains:
\begin{align}
	I =  \frac{1}{N} \int_{v_{min}}^{v_{esc}+v_E} dv\ \frac{d\sigma(v,q)}{d{q}^2} v^2 (I_{MB} - I_{S}),
\end{align}
where,
\begin{align}
    I_{MB}&=\frac{\pi v_0^2}{v_E} 
    \begin{cases}
        g(v-v_E)-g(v+v_E), & v<v_{low}\\
        g(v-v_E)-g(v_{esc}), & otherwise 
    \end{cases},\\
	I_{S}&=2\pi g(v_{esc})
	\begin{cases}
	    2v, & v<v_{low} \\
	    [v_{esc}^2-(v-v_E)^2]/(2v_E), & otherwise
	\end{cases}.
\end{align}
with $v_{low}=v_{esc}-v_E$. $g(x)$ is a one-dimensional Gaussian form:
\begin{align}
	g(v) = e^{-v^2/v_0^2}.
\end{align}
The normalization factor is the same as previously derived \cite{LEWIN199687, PhysRevD.82.075004, PhysRevD.82.023530}:
\begin{align}
    N_{sesc} = \pi^{3/2}v_0^3 \left [\erf(z) - \frac{2}{\sqrt{\pi}}z \left (1+\frac{2}{3}z^2 \right )e^{-z^2} \right ],
\end{align}
with $z=v_{esc}/v_0$.
$I$ is a one-dimensional definite integral. We evaluate it with Gauss-Legendre quadrature.

The limits of the integral have physical constraints. The minimum speed is defined by the minimum recoil energy of a WIMP-nucleus collision at a momentum transfer $q$: $v_{min} = q/(2\mu_T)$, where $\mu_T$ is the reduced mass of the WIMP-nucleus system.  In some approximations the upper limit is simply set to infinity. (Numerically, we approximate $\infty \approx 12 \times v_{0}$.) One can do slightly better by taking the maximum speed to be the galactic escape velocity: $v_{max}=v_{escape} \approx 550\ \mathrm{km/s}$.

\subsection{Quadrature routines}

For the numerical quadrature, we include a third-party function library from \cite{Davis1984}, which we have modified to be compatible with OpenMP parallelization. This routine integrates single-variable functions using an adaptive eight-point Gauss-Legendre algorithm.

As a secondary option, the user can select to use a non-adaptive $n$-th order Gauss-Legendre routine. This can be chosen using the \texttt{quadtype} control word and by setting the order with \texttt{gaussorder}.

It would be straightforward to modify the code to use other quadrature routines, if desired, though special care would need to be taken to ensure thread-safe execution. 

\subsection{Wigner 3-$j$, 6-$j$, 9-$j$ symbols}

We implement a standard set of functions and subroutines for computing the vector-coupling 3-$j$, 6-$j$, and 9-$j$ symbols using the Racah algebraic expressions \cite{edmonds1996angular}. The exact implementation details can be found in the manual.

One method we use to improve  compute time  is to cache Wigner 3-$j$ and 6-$j$ symbols~\cite{edmonds1996angular} (used to evaluate electro-weak matrix elements) in memory at the start of run-time. As a side effect, our tests show that this adds a constant compute time to any given calculation of roughly 0.3 seconds in serial execution and uses roughly 39 MB of memory (for the default table size). As a point of comparison, the $^{131}$Xe example with all-nonzero EFT coefficients in Table \ref{tab:timing} has a run-time of 30 seconds in parallel execution. If we disable the table caching, the run-time is roughly 150 seconds, 5 times longer. The size of the table stored in memory can be controlled via the control file with the keywords \texttt{sj2tablemin} and \texttt{sj2tablemax}.

\subsection{Confluent hypergeometric functions}

For computation of confluent hypergeometric functions (and the required gamma function), we include a third-party function library from \cite{zhang1996computation}, which we have modified to use the intrinsic \texttt{Gamma} function available in Fortran 2008 and later. These functions are used to compute the Bessel spherical and vector harmonics.

\section{Validation and uncertainty}\label{validate}

We validated our Fortran program \texttt{dmscatter} against the Mathematica script \texttt{dmformfactor} (version 6.0). We used $^{29}$Si as a validation case since we have access to the same nuclear density matrix file provided in the \texttt{dmformfactor} package. With a $J=1/2$ ground state, $^{29}$Si also has non-zero coupling to all 15 operators. 

\begin{figure}
    \centering
    \includegraphics[width=.9\columnwidth]{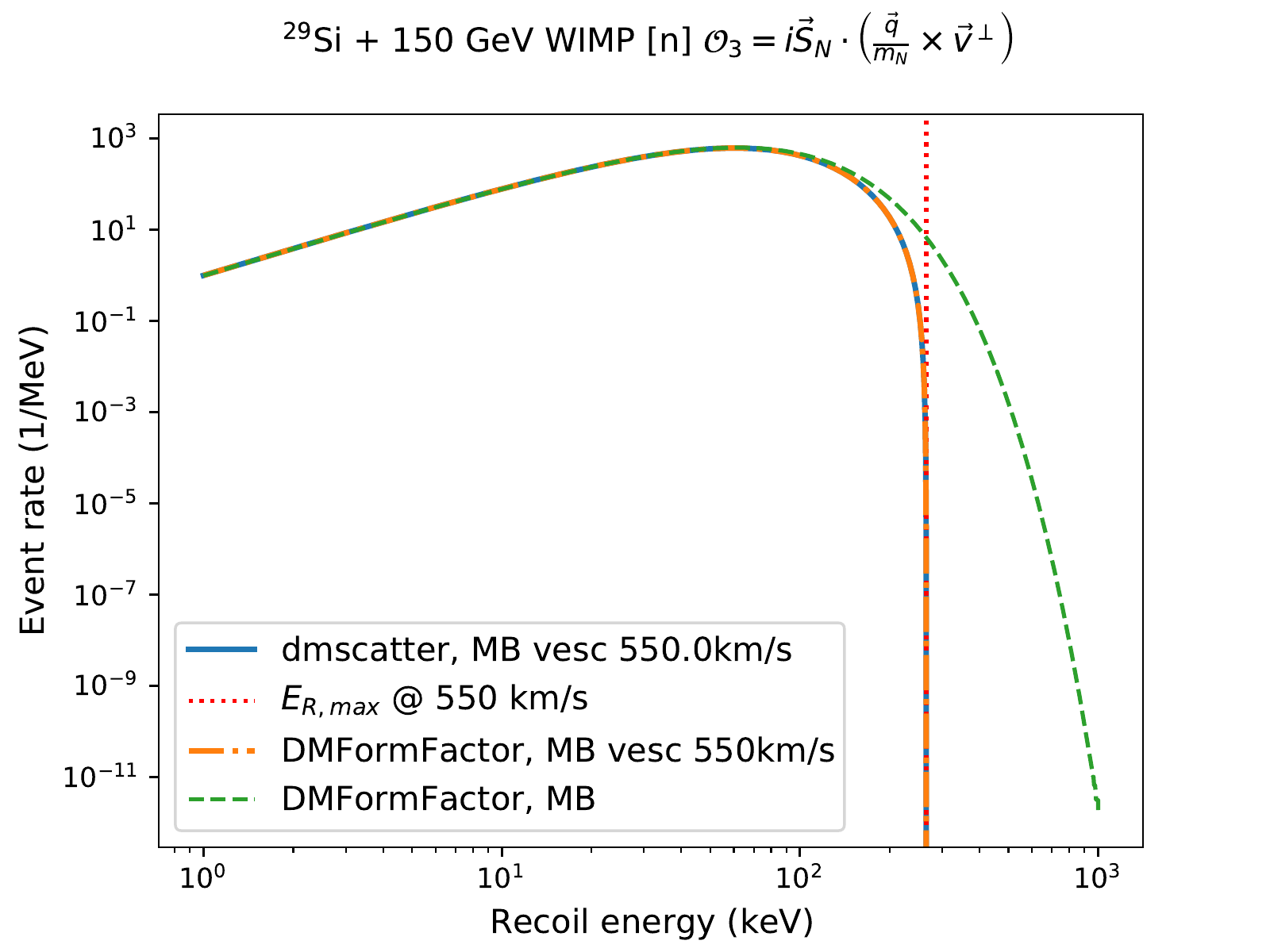}
    \caption{Example differential event rate spectra for $c_3^n=1$ with 150 GeV WIMPS with a $550$ km/s escape velocity on $^{29}$Si computed with Fortran code {\tt dmscatter} (solid blue line) and with Mathematica script {\tt dmformfactor} (orange dash-dot). The green dashed line shows the limit where $v_{esc}\to\infty$.}
    \label{fig: eventrate}
\end{figure}

\begin{figure}
    \centering
    \includegraphics[width=.9\columnwidth]{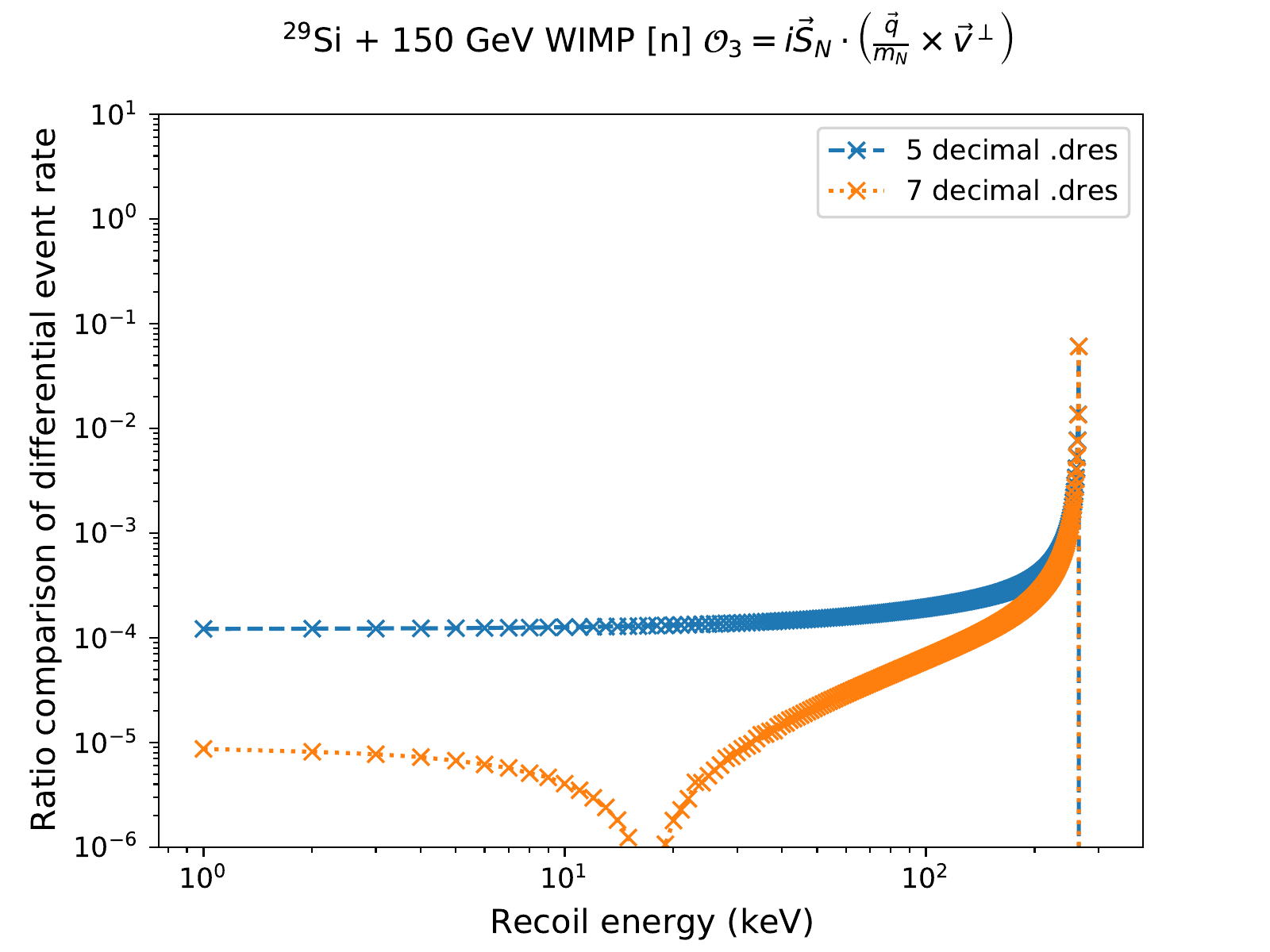}
    \caption{Relative error of {\tt dmscatter} with respect to \texttt{dmformfactor} for the same example. The spike near recoil energy $10^3$ keV is due to round-off error on zero; the event rate there is nearly zero. The (blue) dashed line was calculated with the {\sc Bigstick}-standard 5 decimals of precision in the nuclear structure input (.dres file). The (orange) dotted line has 7 decimal places of precision, matching that used in the {\tt dmformfactor} calculation.}
    \label{fig: error}
\end{figure}

We evaluated the differential event rate for recoil energies from 1 keV to 1000 keV in 1 keV increments for each coefficient individually $c^x_i$; for $i=1,3,4,5,...,15$;  for linearly independent coupling pairs $c_a\cdot c_b=1$ for $(a,b)=$ (1,2), (1,3), (2,3), (4, 5), (5,6), (8,9), (11,12), (11,15), (12,15), and for both $x=p,n$. An example is shown in Figure \ref{fig: eventrate}. In each case we reproduce the results of {\tt dmformfactor}. Typical `error' with respect to \texttt{dmformfactor} is shown in Figure \ref{fig: error}.

There are two sources of error in our calculation. The first is from the model uncertainty of the  nuclear wave functions. This source of error is therefore also present in \texttt{dmformfactor}. While phenomenological calculations can get energies within a few hundred keV~\cite{PhysRevC.74.034315}, other observables often require significant renormalization of operators to agree with experiment, see, e.g.,~\cite{PhysRevC.78.064302}. These errors in observables can have complex origins, arising both from truncations of the model space and higher-order corrections to the corresponding operators~\cite{gysbers2019discrepancy}. Errors in the numerical methods to solve the many-body problem given a set of input parameters are, by comparison, vanishingly small.  Nonetheless experience suggest that in most cases the renormalization is of order one. We have qualitative support of this fact from comparisons of event rate spectra from \textit{ab initio} calculations with increasing model space dimension (scaling with $N_{max}$, the maximum  excitation in an non-interacting harmonic oscillator basis, sometimes written as $N_{max} \, \hbar \omega$ excitations), and from different chiral effective-field theoretical interactions.

The second source of error is from the numerical integration. By comparison, the error from the numerical integration is expected to be many orders of magnitude smaller. Using an adaptive quadrature routine from a standard source \cite{Davis1984}, the code iteratively increases the complexity of the estimator until it achieves a desired relative uncertainty.

\onecolumn

\section{Control words}\label{sec: cwords}
\begin{longtable}{l l p{3.1in} l l}
Keyword & Symbol & Meaning & Units & Default \\
\hline
 dmdens  & $\rho_\chi$ & Local dark matter density. & GeV/cm$^3$ & 0.3 \\
 dmspin  & $j_\chi$    & Instrinsic spin of WIMP particles. & $\hbar$ & $\frac{1}{2}$ \\
 fillnuclearcore & & Logical flag (enter 0 for False, 1 for True) to fill the inert-core single-particle orbitals in the nuclear level densities. Phenomenological shell model calculations typically provide only the density matrices for the active valence-space orbitals. This option automatically assigns these empty matrix elements assuming a totally filled core. & & 1 (true) \\
 gaussorder & & Order of the Gauss-Legendre quadrature to use when using Type 2 quadrature. (See quadtype.) An n-th order routine will perform n function evaluations.  Naturally, a higher order will result in higher precision, but longer compute time. & & 12 \\
 hofrequency & $\hbar \omega$ & Set the harmonic oscillator length by specifying the harmonic oscillator frequency. (b = 6.43/sqrt($\hbar\omega$)). If using an \textit{ab initio} interaction, $\hbar \omega$ should be set to match the value used in the interaction. & MeV & See hoparameter. \\
 hoparameter & $b$ & Harmonic oscillator length. Determines the scale of the nuclear wavefunction interaction. & fm & See eqn. (\ref{bfm}). \\
  maxwellv0 & $v_0$ & Maxwell-Boltzman velocity distribution scaling factor. & km/s & 220.0 \\
  mnucleon & $m_N$ & Mass of a nucleon. It's assumed that $m_p\approx m_n$. & GeV & 0.938272 \\
  ntscale & $N_t$ & Effective number of target nuclei scaling factor. The differential event rate is multiplied by this constant in units of kilogram-days. For example, if the detector had a total effective exposure of 2500 kg days, one might enter 2500 for this value. & kg days & 1.0 \\
 printdensities & & Option to print the nuclear one-body density matrices to screen. & & 0 (false) \\
 pnresponse & & Option to print the nuclear response functions in terms of proton-neutron coupling instead of isospin coupling & & 0 (false) \\
  quadrelerr &  & Desired relative error for the adaptive numerical quadrature routine (quadtype 1).  & & $10^{-6}$ \\
  quadtype & & Option for type of numerical quadrature. (Type 1 = adaptive 8th order Gauss-Legendre quadrature.  Type 2 = static n-th order Gauss-Legendre quadrature.) && 1 (type 1) \\
  sj2tablemax & & Maximum value of $2\times J$ used when caching Wigner 3-$j$ and 6-$j$ functions into memory. & & 12 \\
  sj2tablemin & & Minimum value of $2\times J$ used when caching Wigner 3-$j$ and 6-$j$ functions into memory. & & -2 \\
  useenergyfile & & Logical flag (enter 0 for False, 1 for True) to read energy grid used for calculation from a user-provided file intead of specifying a range. & & 0 (false) \\
  usemomentum & & Logical flag (enter 0 for False, 1 for True) to use momentum transfer intead of recoil energy as the independent variable. & &0 (false) \\
  vearth & $v_{earth}$ & Speed of the earth in the galactic frame. & km/s & 232.0 \\
  vescape & $v_{escape}$ & Galactic escape velocity. Particles moving faster than this speed will escape the galaxy, thus setting an upper limit on the WIMP velocity distribution. & km/s & 12 $\times\ v_{scale}$ \\
  weakmscale & $m_v$ & Weak interaction mass scale. User defined EFT coefficients are divided by $m_v^2$. & GeV & 246.2 \\
  wimpmass & $m_\chi$ & WIMP particle mass. & GeV & 50.0 
\end{longtable}

\end{document}